\documentclass[%
 prl,
 twocolumn,
superscriptaddress,
 amsmath,amssymb,
 aps,
]{revtex4-2}

\usepackage{graphicx}
\usepackage{dcolumn}
\usepackage{bm,bbm}
\usepackage[colorlinks]{hyperref}
\usepackage{xcolor}
\usepackage{xfrac}
\pdfoutput=1

\begin{document}


\title{Non-affinity of liquid networks and bicontinuous mesophases}

\author{Michael S.~Dimitriyev}
 \email{msdim@tamu.edu}
\affiliation{%
 Department of Polymer Science and Engineering, University of Massachusetts, Amherst, MA 01003, USA
}
\affiliation{%
 Department of Materials Science and Engineering, Texas A\&M University, College Station, TX 77843
}

\author{Xueyan Feng}
\affiliation{%
 Department of Macromolecular Science, State Key Laboratory of Molecular Engineering of Polymers, Fudan University, Shanghai, China 200437
}

\author{Edwin L.~Thomas}
\affiliation{%
 Department of Materials Science and Engineering, Texas A\&M University, College Station, TX 77843
}

\author{Gregory M.~Grason}%
 \email{grason@umass.edu}
\affiliation{%
 Department of Polymer Science and Engineering, University of Massachusetts, Amherst, MA 01003, USA
}

\date{\today}


\begin{abstract} 
Amphiphiles self-assemble into a variety of bicontinuous mesophases whose equilibrium structures take the form of high-symmetry cubic networks.  Here, we show that the symmetry-breaking distortions in these systems give rise to anomalously large, non-affine collective deformations, which we argue to be a generic consequence of {\it mass equilibration} within deformed networks.  We propose and study a minimal {\it liquid network} model of bicontinuous networks, in which acubic distortions are modeled by the relaxation of residually-stressed mechanical networks with constant-tension bonds.   We show that non-affinity is strongly dependent on the valency of the network as well as the degree of strain-softening/stiffening force in the bonds.  
Taking diblock copolymer melts as a model system, liquid network theory captures quantitative features of two bicontinuous phases based on comparison with self-consistent field theory predictions and direct experimental characterization of acubic distortions, which are likely to be pronounced in soft amphiphilic systems more generally.

\end{abstract}

\maketitle

The self-assembly of amphiphiles into mesophases with long-range order is of fundamental importance to the formation of structure in biology and nanotechnology.
Beyond the classical lamellar, columnar, and sphere mesophases, there are a variety of bicontinuous mesophases, from amorphous sponges to triply-periodic double-networks.
Of the latter, the double gyroid (DG), double diamond (DD), and double primitive (DP), closely related to triply-periodic (cubic) minimal surfaces, are most commonly observed~\cite{Scriven1976,Thomas1988,Isrealachvili2011}. 
The combination of highly-symmetric morphologies and complex topologies of these structures underlie valuable functional properties, from structural coloration in living organisms~\cite{Galusha2008,Saranathan2010,Saranathan2021} to photonic~\cite{Hur2011,Lee2014} and plasmonic metamaterials~\cite{Mistark2009,Vignolini2012}.

\begin{figure}[h!]
    \centering
    \includegraphics[width=.925\columnwidth]{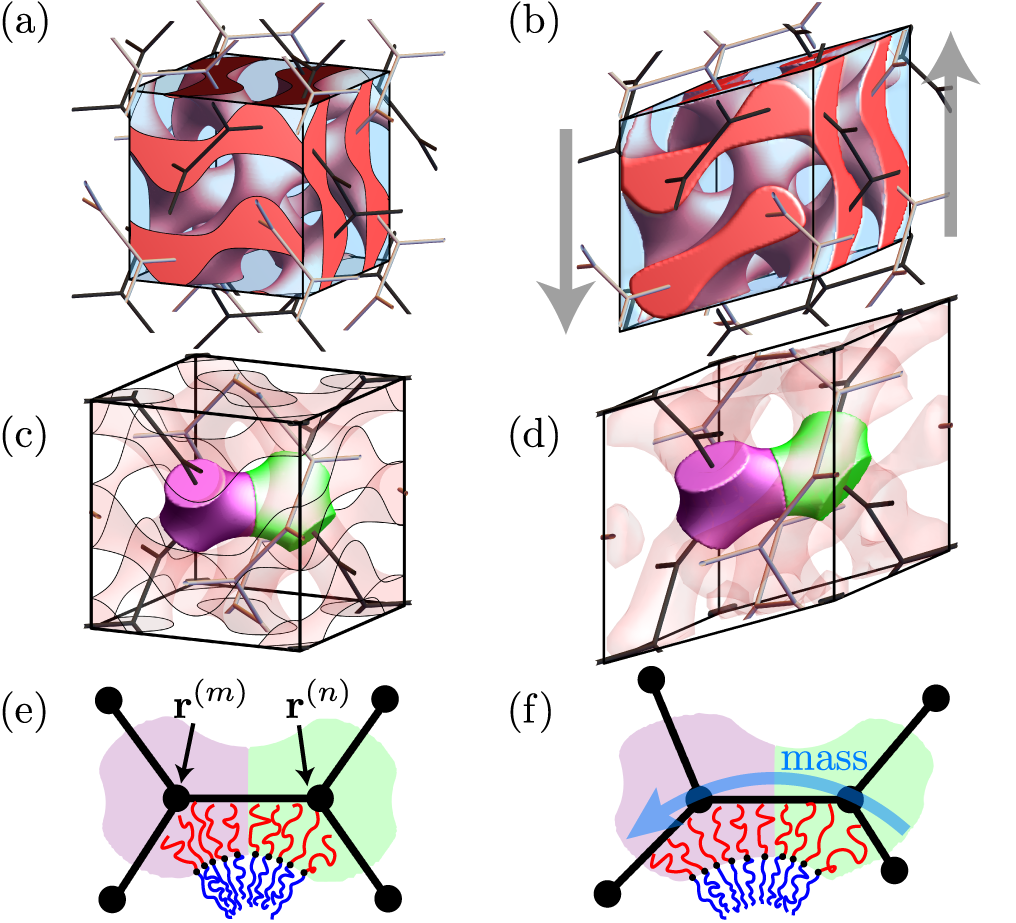}
    \caption{\label{fig:1} Unit cells of a double gyroid (DG) composed of tubular domains (red) separated by a continuous matrix domain (blue) and disjoint skeletal networks from SCF calculations, (a) with full cubic symmetry and (b) after a simple shear. (c),(d) Neighboring ``mesoatomic units'' highlighted in magenta and green, before and after the simple shear, respectively. Schematic arrangement of diblock copolymers with mesoatom pairs, at nodal positions $\mathbf{r}^{(m)}$ and $\mathbf{r}^{(n)}$, possessing equal chain numbers in for cubic structures (e) and net transfer of mass upon symmetry-breaking shear (f).}
\end{figure}

In contrast to atomic or molecular crystals, supramolecular crystals, such as the bicontinuous networks, contain upwards of \mbox{$\sim10^{4}$} macromolecules per unit cell, allowing for more complex distortions.  As shown in Fig.~\ref{fig:1}(a), bicontinuous networks are typically visualized as interconnected tubular domains, separated by slab-like matrices or membranes~\cite{Hyde1997_ch4}.  It is useful to decompose these structures into ``mesoatomic'' units that represent collective groupings centered on the nodal interconnections:  DG, DD, and DP mesoatoms are, respectively, 3-, 4-, and 6-valent~\cite{Reddy2021,GrasonThomas2023}.
Further abstraction of the network domains yields a \textit{skeletal graph} representation (see Fig.~\ref{fig:1}(c)) with nodes centered within the mesoatoms and ``struts'' joining each node.
Recent observations of DG~\cite{Feng2019} and DD~\cite{Feng2023} crystals assembled by block copolymers --- subject to endemic sources of symmetry-breaking stresses (e.g.~due to anisotropic solvent processing and inadequate thermal relaxation via sample annealing) --- exhibit strong, coherent distortions of the otherwise cubic unit cells (Fig.~\ref{fig:1}(b)). 
Experimental reconstructions show a combination of anomalously large fluctuations in strut lengths, yet surprisingly low deviations in bond angles.
This observation hints at the exotic mechanics underlying network relaxation, characterized by pronounced \textit{non-affine} response.
While rigid, amorphous soft materials, such as biopolymer networks, proteins, foams, and gels, have well-characterized non-affinity~\cite{Durian1995,Langer1997,Head2003,DiDonna2005,Conti2009,Broedersz2011,Basu2011,Huisman2011,Broedersz2014,Feng2016,Prakaschchand2020,Pensalfini2023}, and non-affine fluctuations have been studied in crystalline solids~\cite{Jaric1988,Ganguly2013,Das2015}, the mechanism behind the non-affine response in supramolecular networks --- structured fluids with crystalline order --- is apparently fundamentally distinct. Notably, non-affine and symmetry-breaking distortions are thought to be important for the self-assembly of Weyl metamaterials~\cite{Fruchart2018,Jo2021,Park2022}.

In this Letter, we explore the relaxation of supramolecular networks in response to broken cubic symmetry, through the lens of \textit{liquid network theory} (LNT) as a proxy for collective intra-network mass redistribution (Fig.~\ref{fig:1}(b)).  Constant tension networks exhibit highly non-affine response, which is a strong function of nodal valence and symmetry with a linear response encoded in two cubic invariants coupled to simple- and elongational-shear.  LNT predicts trihedral~\footnote{We opt to describe nodes as polyhedra, where each strut emerging from the node is represented by a face. DG nodes are described as trihedral; DD nodes are tetrahedral; DP nodes are hexahedral (described by octahedral symmetry group $O_h$).} DG (the 10-3a net~\cite{Wells1977}) to be generically more non-affine than tetrahedral DD (the 6-4 net), a result that we directly confirm using a combination of self-consistent field (SCF) theory and experimental tomographic reconstructions of acubic network phases of block copolymers.  More generally, we connect this regime of pronounced non-affine relaxation in perfectly-ordered networks to presence of {\it strain-softening} central forces between nodes.

Skeletal network degrees of freedom are specified by node positions $\{\mathbf{r}^{(n)}\}$, along with strut-sharing, nearest-neighbor node pairs $\langle m n\rangle$.
We restrict our attention to network deformations that maintain the number of nodes and the connectivity of the graph.
We take these node positions as the effective, coarse-grained degrees of freedom and consider the simplest form of the free energy, $H$, that can account for intra-nodal mass exchange, composed of nearest-neighbor interactions of the form $H = \sum_{\langle m n\rangle}h(\ell^{(mn)})$, where $\ell^{(mn)} \equiv |\mathbf{r}^{(n)} - \mathbf{r}^{(m)}|$ is the length of the strut joining node $m$ to node $n$ and $h$ is an energy per strut.
The equilibrium condition then requires a balance of forces at each node, i.e.~$\sum_{n \in \langle m n\rangle} \tau(\ell^{(mn)}) \hat{\mathbf{r}}^{(mn)} = 0$, where $\hat{\mathbf{r}}^{(mn)}$ is the unit vector joining node $n$ to node $m$, and $\tau(\ell) \equiv {\rm d}h/{\rm d}\ell$ is the corresponding tension along the strut.

Generic equilibrium conditions for macromolecular assemblies involve a balance of enthalpic costs due to maintaining an intermaterial dividing surface (IMDS) and entropic costs associated with molecular configurations (e.g.~reductions in entropy due to stretching polymers)~\cite{Anderson1988,Matsen1996}.
As depicted in Fig.~\ref{fig:1}(e), macromolecules typically orient normal to the struts; restoring forces due to changes in molecular conformation are thus transverse to the strut.
However, macromolecules can easily translate along an IMDS and even hop between distinct IMDSs (albeit with a free energy barrier); as shown in Fig.~\ref{fig:1}(f), deformations can result in easy intra-network transfer of mass between mesoatoms.
Associating each macromolecule to the nearest strut and assuming the areal density of molecules passing through the IMDS to be approximately constant, the mass per strut is roughly proportional to the strut length $\ell$.
Consequently, to maintain chemical equilibrium in the network, $\tau$ is length-independent ($\tau(\ell) = {\rm const.}$); this is the constitutive relationship for LNT.  Notably, length-minimizing triply-periodic networks with fixed valency are known to correspond to the cubic skeletons of DG, DD and DP~\cite{Alex2019,Alex2023}.

\begin{figure*}[t]
    \centering
    \includegraphics[width=.9\textwidth]{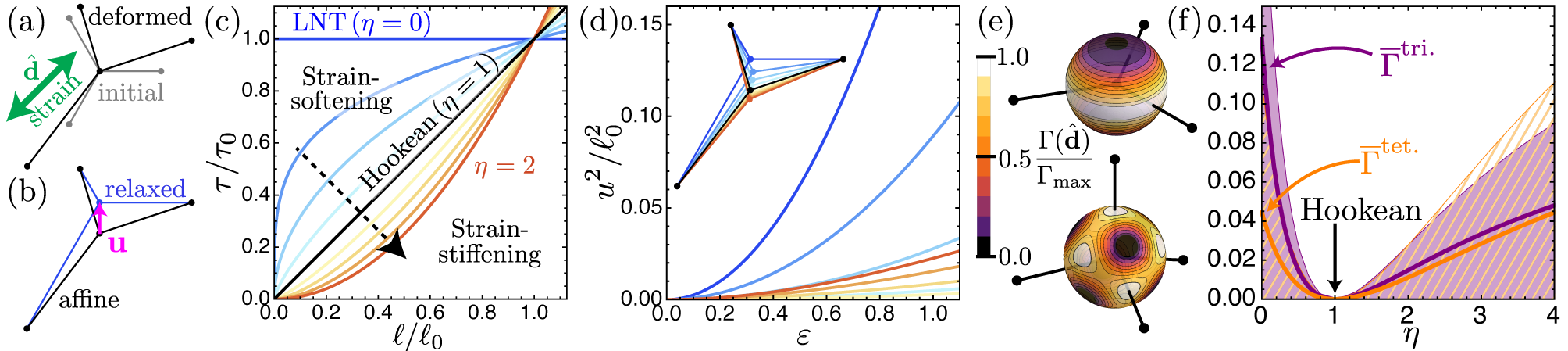}
    \caption{\label{fig:2} (a) 3-strut planar graph with initial strut lengths $\ell_0$ and angle of $120^\circ$ between neighboring struts, superposed with affinely stretched variant (along $\hat{\mathbf{d}}$, in the plane). (b) Non-affine relaxation, characterized by central node displacement $\mathbf{u}$ under equilibrium of LNT. (c) Power-law constitutive relations for tension $\tau(\ell)$ as a function of length $\ell$. (d) Non-affine displacement $|{\bf u}|^2/\ell_0^2$, as a function of strain $\varepsilon$. (e) Spherical map of direction-dependent  non-affinity $\Gamma\equiv |{\bf u}|^2/(\ell_0^2\varepsilon^2)$ for a trihedral (top) and a tetrahedral (bottom) graph. (f) Average non-affine response for trihedral and tetrahedral graphs as a function of $\eta$. Shaded regions represent the full range of $\Gamma(\hat{\mathbf{d}})$.}
\end{figure*}

To explore the behavior of LNT, we first consider the planar 3-strut graph shown in Fig.~\ref{fig:2}(a), whose initial configuration of struts with equal length $\ell_0$ and angle of $120^{\circ}$ results in mechanical equilibrium for the central node for any constitutive form of $\tau(\ell)$.
The boundary nodes are then {\it affinely displaced} by stretching along the direction $\hat{\mathbf{d}}$ by a factor $\varepsilon$; the deformation matrix is $\Lambda_{ij} = \delta_{ij} + \varepsilon \hat{d}_i\hat{d}_j$.
Finally, relaxation of the central node, while holding boundary nodes fixed, results in a displacement $\mathbf{u}$, as depicted in Fig.~\ref{fig:2}(b); a non-zero value of $\mathbf{u}$ indicates a \emph{non-affine deformation} of the graph.
For LNT (constant tension), equilibrium requires strut angles maintain $120^\circ$-coordination, which generally necessitates a non-affine displacement of the central node.
This can be regarded as a 1D version of Plateau's laws for soap films~\cite{Isenberg1992,Ivanov1994MinimalNT}.
Beyond trihedral graphs, equilibrium conditions for LNT require strut orientations to sum to zero at each node.
Thus, nodes exhibiting tetrahedral and hexahedral symmetry, such as those seen in DD and DP, are LNT equilibria, as are the more exotic pentahedral nodes seen in DD mirror twin boundaries~\cite{Feng2023}.
More generally, we can consider a broader class of power-law constitutive relations
\begin{equation}\label{eq:constitutive_relation}
    \tau(\ell) = \tau_0 \left(\frac{\ell}{\ell_0}\right)^{\eta} \, ,
\end{equation}
where $\tau_0$ is the residual tension of struts in the initial configuration.
The exponent $\eta$ controls the character of the mechanical response, as shown in Fig.~\ref{fig:2}(c): for $\eta = 1$, the struts obey Hooke's law; for $0 < \eta < 1$, the network {\it strain-softens}; for $\eta > 1$, the network {\it strain-stiffens}; $\eta = 0$ corresponds to LNT (i.e.~constant tension).
The increasing magnitude of non-affine deformation $|{\bf u}|^2$ with strain $\varepsilon$ varies with $\eta$, as shown in Fig.~\ref{fig:2}(c).
This measure of the non-affine response increases quadratically with strain, $|{\bf u}|^2/\ell_0^2 = \Gamma\varepsilon^2$, as is generally expected for mechanical networks for $\varepsilon \ll 1$, where $\Gamma$ is the \emph{non-affinity} parameter~\cite{DiDonna2005,Wyart2008,Basu2011,Broedersz2011,Carrillo2013}.
We find that LNT is characterized by especially large values of $\Gamma$, compared to strain-stiffening forces that are typical of many fiber networks~\cite{Gardel2004,Storm2005,Lieleg2007,Broedersz2014,Sharma2016}. 
As shown in the SI, similar values can be achieved in finitely-extensible flexible and semi-flexible polymer networks~\cite{Marko1995,Rubinstein2003}, but only in the highly non-Hookean, high-extension limit.
Moreover, the volume constraint ensures that struts maintain non-zero tension and places LNT in a similar class of residually-stressed mechanical networks as those described by the classical description of rubber elasticity~\cite{Treloar1975,Flory1985}, involving Hookean networks with affine response, independent of network geometry~\cite{Alexander1998}.

Next, consider trihedral and tetrahedral graphs in 3D, representing a nodal region in DG and DD, respectively.
The resulting non-affinity parameter $\Gamma(\hat{\mathbf{d}})$ has a directional dependence that is plotted on the unit sphere, superimposed on both graphs in Fig.~\ref{fig:2}(e); by symmetry of the uniaxial deformation, $\Gamma(-\hat{\mathbf{d}}) = \Gamma(\hat{\mathbf{d}})$.
The restoring force from each strut is proportional to the projection of the strut direction onto the stretch direction $\hat{\mathbf{d}}$; consequently, the DP hexahedral nodes always maintain affine deformations.
The non-affine response is therefore maximal along the strut directions for both graphs and is zero along the 3-fold axis (out-of-plane normal) of the trihedral graph and along each of the 4-fold rotoinversion axes of the tetrahedral graph.
The full range of the non-affinity parameter for both graphs is shown in Fig.~\ref{fig:2}(f).
We see that LNT has the largest non-affine response for both networks over a broad range of non-negative values of $\eta$, with larger values for the tetrahedral graph when $\eta > 4$; as $\eta \to \infty$, the maximal values of $\Gamma^{\rm tri.}$ and $\Gamma^{\rm tet.}$ approach $1/4$ and $4/9$, respectively.
Moreover, the trihedral graph exhibits a consistently larger maximum of the non-affinity parameter for small $\eta$; the tetrahedral graph becomes more non-affine than the trihedral for $\eta > 2$.
Since networks may contain trihedral or tetrahedral nodes (or both for (3,4) nets~\cite{Wells1977} and certain Fischer-Koch structures~\cite{Fischer1996}) with a variety of orientations, it is useful to quantify the averaged response for both types of nodes, which we obtain by averaging $\Gamma(\hat{\mathbf{d}})$ over all possible stretching directions on the unit sphere.
The resulting averaged non-affinity, $\overline{\Gamma}$, is plotted in Fig.~\ref{fig:2}(f); note that $\overline{\Gamma}^{\rm tri.}/\overline{\Gamma}^{\rm tet.} = 3$ (see SI).

Next, we turn to the linear response of full DG and DD networks at fixed volume; calculation details are shown in the SI.
Given a general deformation matrix $\bm{\Lambda}$, the non-affine response is a function of the strain tensor, \mbox{$\bm{\varepsilon} = \left(\bm{\Lambda}^T\bm{\Lambda}-\mathbbm{1}\right)/2$}, which is, by construction, symmetric and invariant under rigid rotations of the network~\cite{LandauLifshitz1986}.
The non-affine response can be characterized by $\langle | {\bf u}|^2 \rangle$, where the average is taken over all nodes in a unit cell.
In the linear response regime, the quadratic scaling with $\varepsilon_{ij}$ is characterized by a \emph{non-affinity tensor} $\Gamma^{ijkl}$ and takes on the form
\begin{equation}
\frac{\langle |\mathbf{u}|^2 \rangle}{\ell_0^2} = \Gamma^{ijkl} \varepsilon_{ij}\varepsilon_{kl} = \Gamma_{\rm ext}\,\sum_i \frac{\hat{\varepsilon}_{ii}^2}{2} + \Gamma_{\rm shear}\sum_{i\neq j} \hat{\varepsilon}_{ij}^2\, ,
\end{equation}
where $\hat{\varepsilon}_{ij} \equiv \varepsilon_{ij} - \varepsilon_{kk}\delta_{ij}/3$ is the deviatoric (i.e.~traceless) strain.
For the cubic networks considered here, we can express the tensor in terms of two independent components, $\Gamma_{\rm ext}$ and $\Gamma_{\rm shear}$, representing the response to volume-preserving extensional and shear strains, respectively (see SI).

The resulting extensional and shear components of the non-affinity tensor are given in Table~\ref{tab:nonaffinity} for general values of $\eta$.
LNT predicts a ratio between the non-affine response to shear deformation $\Gamma^{\rm DG}_{\rm shear}/\Gamma^{\rm DD}_{\rm shear} = 3$ that is in agreement with average non-affinity ratio of the trihedral and tetrahedral nodes.
Because the extensional strain involves compression and elongation along the $[100]$ family of directions relative to the unit cell, which align with the rotoinversion axes of the tetrahedral nodes, $\Gamma_{\rm ext}^{\rm DD}$ is uniformly zero for all values of $\eta$.
Notably, for general $\eta > 0$, $\Gamma_{\rm ext}^{\rm DG}$ becomes smaller than $\Gamma_{\rm shear}^{\rm DG}$.

\begin{table}[h]
\renewcommand*{\arraystretch}{1.4}
\begin{ruledtabular}
\begin{tabular}{l|cccc}
 & $\Gamma^{\rm DG}_{\rm ext}$ & $\Gamma^{\rm DG}_{\rm shear}$ & $\Gamma^{\rm DD}_{\rm ext}$ & $\Gamma^{\rm DD}_{\rm shear}$ \\
 \hline\\[-10pt]
gen.~$\eta$ & $\frac{3}{4}\left(\frac{1-\eta}{1 + 3\eta}\right)^2$ & $\frac{1}{8}\left(\frac{1-\eta}{1 + \eta}\right)^2$ & 0 & $\frac{1}{6}\left(\frac{1 - \eta}{2 + \eta}\right)^2$\\
LNT & $0.75$ & $0.125$ & 0 & $\approx 0.042$ \\
SCF & $0.20_{\pm 0.02}$ & $0.17_{\pm 0.02}$ & $0.02_{\pm 0.01}$ & $0.28_{\pm 0.02}$ \\
\end{tabular}
\end{ruledtabular}
\caption{\label{tab:nonaffinity} Components of the cubic non-affinity tensor $\Gamma$ for DG and DD, for power-law strut tension $\tau(\ell)\propto \ell^\eta$, liquid network theory (LNT), and self-consistent field (SCF) theory linear diblock copolymer melts.}
\end{table}

\begin{figure}[t]
    \centering
    \includegraphics[width=.9125\columnwidth]{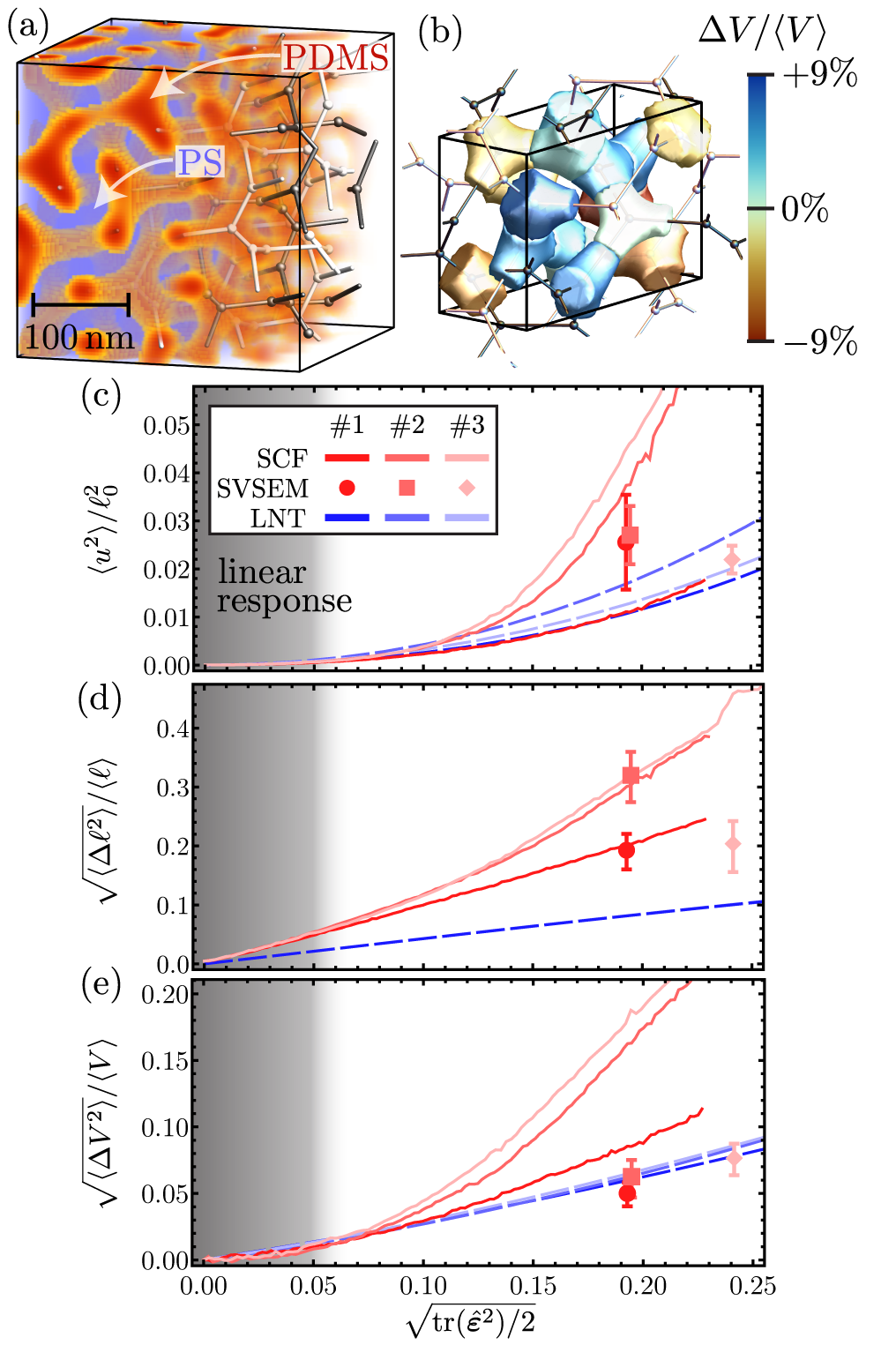}
    \caption{\label{fig:3} (a) 3D tomographic reconstruction from SVSEM data of a DG grain from PS-PDMS thin film, with extracted skeletal graph. (b) Distribution of nodal volumes in a 3D construction of a unit cell in DG sample \#1, showing $\sim \pm 9 \%$ variations in volume. Below, plots show (c) average non-affine displacement $\langle |\mathbf{u}|^2 \rangle$, (d) strut length variations $\sqrt{\langle \Delta \ell^2 \rangle}/\langle \ell \rangle$, and (e) variance in nodal volume $\sqrt{\langle \Delta V^2 \rangle}/\langle V \rangle$, for experiments (markers), simulated deformation paths from SCF calculations (red curves), and corresponding LNT predictions (blue, dashed curves) as a function of total strain $\sqrt{{\rm tr}(\hat{\bm{\varepsilon}}^2)/2}$.}
\end{figure}

Now we consider our results in the context of diblock copolymers (BCPs), based on experiments involving DG and DD networks formed from polystyrene-$b$-polydimethylsiloxane (PS-PDMS), as well as associated self-consistent field (SCF) calculations of equilibrium structures (see SI).
First, we performed a series of SCF computations that explore the equilibrium morphologies formed by BCPs when the cubic symmetry of the DG/DD unit cell is broken by an applied affine transformation $\bm{\Lambda}$.
Using a skeletonization algorithm, we determined the non-affine displacements $\langle |\mathbf{u}|^2 \rangle$ of nodes along separate extensional and shear strain deformation paths (see SI).
This way, we were able to measure non-affinity components from SCF calculations, shown in Table~\ref{tab:nonaffinity}.
For DG, we found that the shear component is comparable to predictions of LNT, but the extensional component is somewhat smaller than expected.
This significant depression of $\Gamma^{\rm DG}_{\rm ext}$ suggests that terms beyond LNT may prove important for certain strain directions; indeed, our mechanical network model suggests that a small, non-zero value of $\eta$ is needed to capture this scale of non-affinity.
While such a fractional tension-length constitutive relationship is unlikely to have a clear physical interpretation, our use of $\eta$ in the power-law form $\tau(\ell)$ given in Eq.~\ref{eq:constitutive_relation} is simply to facilitate the analysis in this study.
As shown in the SI, our results hold under a much more general set of constitutive relationships $\tau(\ell)$, wherein $\eta = \ell \tau'/\tau$, which will generically depend on deformation. 
SCF calculations show a small, non-zero value for $\Gamma^{\rm DD}_{\rm ext}$ as well as a larger-than-expected value of $\Gamma^{\rm DD}_{\rm shear}$.
Such discrepancies suggest that effects beyond the minimal LNT, such as inter-network interactions, may play a larger role in the response of DD.

We analyzed prior experimental data, which consists of 3D density fields extracted using ``slice-and-view'' scanning electron microscopy (SVSEM)~\cite{Feng2019,Reddy2021}.
This technique yields nm-scale voxels representing local density measurements of PDMS relative to PS, resulting in 3D reconstructions, such as shown in Fig.~\ref{fig:3}(a), which then can be skeletonized using a similar algorithm as used for the density fields obtained from SCF calculations~\cite{Prasad2018}.
The samples include structure sampled from three distinct grains of a DG polycrystal and two distinct grains of a DD polycrystal.
While each of the grains possess distinct triclinically deformed unit cells (obtained from Fourier analysis of the density field data), we can compare against predictions of LNT and molecular SCF theory by simulating constant-volume deformations paths that include the triclinic unit cells for each of the grains (i.e.~maintaining roughly constant ratios of shear to extensional strain).
The average non-affine displacement along each family of deformations for the DG samples is shown in Fig.~\ref{fig:3}(b) for SCF calculations and LNT, along with experimental measurements; note that the error bars come from unit-cell averaged measurement of non-affine displacement from multiple unit cells within each 3D reconstruction.
We plot these distinct deformation paths, which combine mixtures of extensional and shear strain, as functions of a common strain measure $\sqrt{{\rm tr}(\hat{\bm{\varepsilon}}^2)/2}$, a combination of extensional and shear strain (see SI).
While there is good agreement between LNT and SCF for low strain, SCF results depart from LNT predictions for strain values on the order of $5-10\%$.
For larger strains, the non-affine response attains significant non-quadratic contributions involving higher order couplings between extensional and shear strain that fall well outside of the linear response regime.
Nevertheless, we note that the parameter-free LNT does a remarkable job in capturing the magnitude of non-affinity for the experimental DG samples, which evidently exhibit significant acubic distortion ($\approx 20\%$).
Further studies involving careful processing and annealing are needed to fully relax the networks and may result in better agreement between experiment and theory.

Finally, we address our central hypothesis that mass transport plays a key role in the equilibration of deformed networks.
Since we assumed that macromolecules associate to struts, the variance in strut length, shown in Fig.~\ref{fig:3}(c), is related to fluctuations in the total mass of molecules associated to a given strut. These fluctuations may originate from variations in macromolecule density and conformations at constant mass per strut, and thus involve elastic contributions to the strut tension.
A different measure of mass transport is the variation in the volume per node (or mesoatom), determined by counting the mass contribution per voxel and associating each voxel to the nearest node in space (see SI).
For example, the a single unit cell from an experimental sample shows $\sim 10\%$ variation in volume between nodes (see Fig.~\ref{fig:3}(b)) that are equivalent in the undeformed, cubic configuration.
Applying the same method to the DG grains and simulated deformation paths, we find that variance in nodal volume increases with strain and again, as shown in Fig.~\ref{fig:3}(e), LNT exhibits a smaller variance than SCF calculations as well as experimental measurements.

We have shown that, to a first approximation, the collective response of supramolecular networks to symmetry-breaking deformations is controlled by equilibration of mass along network elements, captured via a simple principle of length-minimization.
While both experiment and SCF theory exhibit near-quantitative and fit free agreement with LNT in the linear response regime, it is clear that additional contributions from macromolecule elasticity as well as packing constraints (and thus intranetwork interactions) are needed, particularly in the large-strain regime.
Nevertheless, LNT naturally explains prior observations of ``rigid'' strut angle relationships and ``soft'' strut length requirements for BCP network deformations and defects~\cite{Feng2019,Feng2023}.
Furthermore, LNT can rationalize an observed ``node-splitting'' transition observed in DD (see SI), particularly under modest extensional strain, where tetrahedral nodes split into pairs of trihedral nodes.
Since LNT equilibria correspond to locally length-minimizing networks, node-splitting events tend towards globally length-minimizing configurations, known as ``Steiner Networks,'' composed of purely trihedral nodes~\cite{Alex2019,Alex2023}.
As such, LNT provides a minimal basis for understanding deformations of supramolecular networks well beyond linear response, into the regime of defects and structural transitions.

The authors thank B.~Greenvall for stimulating discussions and valuable comments on this work. This research was supported by the U.S. Department of Energy (DOE), Office of Basic Energy Sciences, Division of Materials Sciences and Engineering, under award DE-SC0022229. SCF computations were performed on the Unity Cluster at the Massachusetts Green High Performance Computing Center.


\bibliography{refs}

\begin{thebibliography}{55}%
\makeatletter
\providecommand \@ifxundefined [1]{%
 \@ifx{#1\undefined}
}%
\providecommand \@ifnum [1]{%
 \ifnum #1\expandafter \@firstoftwo
 \else \expandafter \@secondoftwo
 \fi
}%
\providecommand \@ifx [1]{%
 \ifx #1\expandafter \@firstoftwo
 \else \expandafter \@secondoftwo
 \fi
}%
\providecommand \natexlab [1]{#1}%
\providecommand \enquote  [1]{``#1''}%
\providecommand \bibnamefont  [1]{#1}%
\providecommand \bibfnamefont [1]{#1}%
\providecommand \citenamefont [1]{#1}%
\providecommand \href@noop [0]{\@secondoftwo}%
\providecommand \href [0]{\begingroup \@sanitize@url \@href}%
\providecommand \@href[1]{\@@startlink{#1}\@@href}%
\providecommand \@@href[1]{\endgroup#1\@@endlink}%
\providecommand \@sanitize@url [0]{\catcode `\\12\catcode `\$12\catcode
  `\&12\catcode `\#12\catcode `\^12\catcode `\_12\catcode `\%12\relax}%
\providecommand \@@startlink[1]{}%
\providecommand \@@endlink[0]{}%
\providecommand \url  [0]{\begingroup\@sanitize@url \@url }%
\providecommand \@url [1]{\endgroup\@href {#1}{\urlprefix }}%
\providecommand \urlprefix  [0]{URL }%
\providecommand \Eprint [0]{\href }%
\providecommand \doibase [0]{https://doi.org/}%
\providecommand \selectlanguage [0]{\@gobble}%
\providecommand \bibinfo  [0]{\@secondoftwo}%
\providecommand \bibfield  [0]{\@secondoftwo}%
\providecommand \translation [1]{[#1]}%
\providecommand \BibitemOpen [0]{}%
\providecommand \bibitemStop [0]{}%
\providecommand \bibitemNoStop [0]{.\EOS\space}%
\providecommand \EOS [0]{\spacefactor3000\relax}%
\providecommand \BibitemShut  [1]{\csname bibitem#1\endcsname}%
\let\auto@bib@innerbib\@empty
\bibitem [{\citenamefont {Scriven}(1976)}]{Scriven1976}%
  \BibitemOpen
  \bibfield  {author} {\bibinfo {author} {\bibfnamefont {L.~E.}\ \bibnamefont
  {Scriven}},\ }\bibfield  {title} {\bibinfo {title} {{Equilibrium bicontinuous
  structure}},\ }\href {https://doi.org/10.1038/263123a0} {\bibfield  {journal}
  {\bibinfo  {journal} {Nature}\ }\textbf {\bibinfo {volume} {263}},\ \bibinfo
  {pages} {123} (\bibinfo {year} {1976})}\BibitemShut {NoStop}%
\bibitem [{\citenamefont {Thomas}\ \emph {et~al.}(1988)\citenamefont {Thomas},
  \citenamefont {Anderson}, \citenamefont {Henkee},\ and\ \citenamefont
  {Hoffman}}]{Thomas1988}%
  \BibitemOpen
  \bibfield  {author} {\bibinfo {author} {\bibfnamefont {E.~L.}\ \bibnamefont
  {Thomas}}, \bibinfo {author} {\bibfnamefont {D.~M.}\ \bibnamefont
  {Anderson}}, \bibinfo {author} {\bibfnamefont {C.~S.}\ \bibnamefont
  {Henkee}},\ and\ \bibinfo {author} {\bibfnamefont {D.}~\bibnamefont
  {Hoffman}},\ }\bibfield  {title} {\bibinfo {title} {{Periodic area-minimizing
  surfaces in block copolymers}},\ }\href {https://doi.org/10.1038/334598a0}
  {\bibfield  {journal} {\bibinfo  {journal} {Nature}\ }\textbf {\bibinfo
  {volume} {334}},\ \bibinfo {pages} {598} (\bibinfo {year}
  {1988})}\BibitemShut {NoStop}%
\bibitem [{\citenamefont {Israelachvili}(2011)}]{Isrealachvili2011}%
  \BibitemOpen
  \bibfield  {author} {\bibinfo {author} {\bibfnamefont {J.~N.}\ \bibnamefont
  {Israelachvili}},\ }\bibfield  {title} {\bibinfo {title} {20 - soft and
  biological structures},\ }in\ \href
  {https://doi.org/https://doi.org/10.1016/B978-0-12-375182-9.10020-X} {\emph
  {\bibinfo {booktitle} {Intermolecular and Surface Forces (Third Edition)}}},\
  \bibinfo {editor} {edited by\ \bibinfo {editor} {\bibfnamefont {J.~N.}\
  \bibnamefont {Israelachvili}}}\ (\bibinfo  {publisher} {Academic Press},\
  \bibinfo {address} {San Diego},\ \bibinfo {year} {2011})\ \bibinfo {edition}
  {third edition}\ ed.,\ pp.\ \bibinfo {pages} {535--576}\BibitemShut {NoStop}%
\bibitem [{\citenamefont {Galusha}\ \emph {et~al.}(2008)\citenamefont
  {Galusha}, \citenamefont {Richey}, \citenamefont {Gardner}, \citenamefont
  {Cha},\ and\ \citenamefont {Bartl}}]{Galusha2008}%
  \BibitemOpen
  \bibfield  {author} {\bibinfo {author} {\bibfnamefont {J.~W.}\ \bibnamefont
  {Galusha}}, \bibinfo {author} {\bibfnamefont {L.~R.}\ \bibnamefont {Richey}},
  \bibinfo {author} {\bibfnamefont {J.~S.}\ \bibnamefont {Gardner}}, \bibinfo
  {author} {\bibfnamefont {J.~N.}\ \bibnamefont {Cha}},\ and\ \bibinfo {author}
  {\bibfnamefont {M.~H.}\ \bibnamefont {Bartl}},\ }\bibfield  {title} {\bibinfo
  {title} {Discovery of a diamond-based photonic crystal structure in beetle
  scales},\ }\href {https://doi.org/10.1103/PhysRevE.77.050904} {\bibfield
  {journal} {\bibinfo  {journal} {Phys. Rev. E}\ }\textbf {\bibinfo {volume}
  {77}},\ \bibinfo {pages} {050904} (\bibinfo {year} {2008})}\BibitemShut
  {NoStop}%
\bibitem [{\citenamefont {Saranathan}\ \emph {et~al.}(2010)\citenamefont
  {Saranathan}, \citenamefont {Osuji}, \citenamefont {Mochrie}, \citenamefont
  {Noh}, \citenamefont {Narayanan}, \citenamefont {Sandy}, \citenamefont
  {Dufresne},\ and\ \citenamefont {Prum}}]{Saranathan2010}%
  \BibitemOpen
  \bibfield  {author} {\bibinfo {author} {\bibfnamefont {V.}~\bibnamefont
  {Saranathan}}, \bibinfo {author} {\bibfnamefont {C.~O.}\ \bibnamefont
  {Osuji}}, \bibinfo {author} {\bibfnamefont {S.~G.~J.}\ \bibnamefont
  {Mochrie}}, \bibinfo {author} {\bibfnamefont {H.}~\bibnamefont {Noh}},
  \bibinfo {author} {\bibfnamefont {S.}~\bibnamefont {Narayanan}}, \bibinfo
  {author} {\bibfnamefont {A.}~\bibnamefont {Sandy}}, \bibinfo {author}
  {\bibfnamefont {E.~R.}\ \bibnamefont {Dufresne}},\ and\ \bibinfo {author}
  {\bibfnamefont {R.~O.}\ \bibnamefont {Prum}},\ }\bibfield  {title} {\bibinfo
  {title} {{Structure, function, and self-assembly of single network gyroid
  (I4132) photonic crystals in butterfly wing scales}},\ }\href
  {https://doi.org/10.1073/pnas.0909616107} {\bibfield  {journal} {\bibinfo
  {journal} {Proceedings of the National Academy of Sciences}\ }\textbf
  {\bibinfo {volume} {107}},\ \bibinfo {pages} {11676} (\bibinfo {year}
  {2010})}\BibitemShut {NoStop}%
\bibitem [{\citenamefont {Saranathan}\ \emph {et~al.}(2021)\citenamefont
  {Saranathan}, \citenamefont {Narayanan}, \citenamefont {Sandy}, \citenamefont
  {Dufresne},\ and\ \citenamefont {Prum}}]{Saranathan2021}%
  \BibitemOpen
  \bibfield  {author} {\bibinfo {author} {\bibfnamefont {V.}~\bibnamefont
  {Saranathan}}, \bibinfo {author} {\bibfnamefont {S.}~\bibnamefont
  {Narayanan}}, \bibinfo {author} {\bibfnamefont {A.}~\bibnamefont {Sandy}},
  \bibinfo {author} {\bibfnamefont {E.~R.}\ \bibnamefont {Dufresne}},\ and\
  \bibinfo {author} {\bibfnamefont {R.~O.}\ \bibnamefont {Prum}},\ }\bibfield
  {title} {\bibinfo {title} {{Evolution of single gyroid photonic crystals in
  bird feathers}},\ }\href {https://doi.org/10.1073/pnas.2101357118} {\bibfield
   {journal} {\bibinfo  {journal} {Proceedings of the National Academy of
  Sciences}\ }\textbf {\bibinfo {volume} {118}},\ \bibinfo {pages}
  {e2101357118} (\bibinfo {year} {2021})}\BibitemShut {NoStop}%
\bibitem [{\citenamefont {Hur}\ \emph {et~al.}(2011)\citenamefont {Hur},
  \citenamefont {Francescato}, \citenamefont {Giannini}, \citenamefont {Maier},
  \citenamefont {Hennig},\ and\ \citenamefont {Wiesner}}]{Hur2011}%
  \BibitemOpen
  \bibfield  {author} {\bibinfo {author} {\bibfnamefont {K.}~\bibnamefont
  {Hur}}, \bibinfo {author} {\bibfnamefont {Y.}~\bibnamefont {Francescato}},
  \bibinfo {author} {\bibfnamefont {V.}~\bibnamefont {Giannini}}, \bibinfo
  {author} {\bibfnamefont {S.~A.}\ \bibnamefont {Maier}}, \bibinfo {author}
  {\bibfnamefont {R.~G.}\ \bibnamefont {Hennig}},\ and\ \bibinfo {author}
  {\bibfnamefont {U.}~\bibnamefont {Wiesner}},\ }\bibfield  {title} {\bibinfo
  {title} {{Three-Dimensionally Isotropic Negative Refractive Index Materials
  from Block Copolymer Self-Assembled Chiral Gyroid Networks}},\ }\href
  {https://doi.org/10.1002/anie.201104888} {\bibfield  {journal} {\bibinfo
  {journal} {Angewandte Chemie International Edition}\ }\textbf {\bibinfo
  {volume} {50}},\ \bibinfo {pages} {11985} (\bibinfo {year}
  {2011})}\BibitemShut {NoStop}%
\bibitem [{\citenamefont {Lee}\ \emph {et~al.}(2014)\citenamefont {Lee},
  \citenamefont {Koh}, \citenamefont {Singer}, \citenamefont {Jeon},
  \citenamefont {Maldovan}, \citenamefont {Stein},\ and\ \citenamefont
  {Thomas}}]{Lee2014}%
  \BibitemOpen
  \bibfield  {author} {\bibinfo {author} {\bibfnamefont {J.-H.}\ \bibnamefont
  {Lee}}, \bibinfo {author} {\bibfnamefont {C.~Y.}\ \bibnamefont {Koh}},
  \bibinfo {author} {\bibfnamefont {J.~P.}\ \bibnamefont {Singer}}, \bibinfo
  {author} {\bibfnamefont {S.-J.}\ \bibnamefont {Jeon}}, \bibinfo {author}
  {\bibfnamefont {M.}~\bibnamefont {Maldovan}}, \bibinfo {author}
  {\bibfnamefont {O.}~\bibnamefont {Stein}},\ and\ \bibinfo {author}
  {\bibfnamefont {E.~L.}\ \bibnamefont {Thomas}},\ }\bibfield  {title}
  {\bibinfo {title} {{25th Anniversary Article: Ordered Polymer Structures for
  the Engineering of Photons and Phonons}},\ }\href
  {https://doi.org/10.1002/adma.201303456} {\bibfield  {journal} {\bibinfo
  {journal} {Advanced Materials}\ }\textbf {\bibinfo {volume} {26}},\ \bibinfo
  {pages} {532} (\bibinfo {year} {2014})}\BibitemShut {NoStop}%
\bibitem [{\citenamefont {Mistark}\ \emph {et~al.}(2009)\citenamefont
  {Mistark}, \citenamefont {Park}, \citenamefont {Yalcin}, \citenamefont {Lee},
  \citenamefont {Yavuzcetin}, \citenamefont {Tuominen}, \citenamefont
  {Russell},\ and\ \citenamefont {Achermann}}]{Mistark2009}%
  \BibitemOpen
  \bibfield  {author} {\bibinfo {author} {\bibfnamefont {P.~A.}\ \bibnamefont
  {Mistark}}, \bibinfo {author} {\bibfnamefont {S.}~\bibnamefont {Park}},
  \bibinfo {author} {\bibfnamefont {S.~E.}\ \bibnamefont {Yalcin}}, \bibinfo
  {author} {\bibfnamefont {D.~H.}\ \bibnamefont {Lee}}, \bibinfo {author}
  {\bibfnamefont {O.}~\bibnamefont {Yavuzcetin}}, \bibinfo {author}
  {\bibfnamefont {M.~T.}\ \bibnamefont {Tuominen}}, \bibinfo {author}
  {\bibfnamefont {T.~P.}\ \bibnamefont {Russell}},\ and\ \bibinfo {author}
  {\bibfnamefont {M.}~\bibnamefont {Achermann}},\ }\bibfield  {title} {\bibinfo
  {title} {Block-copolymer-based plasmonic nanostructures},\ }\href
  {https://doi.org/10.1021/nn901245w} {\bibfield  {journal} {\bibinfo
  {journal} {ACS Nano}\ }\textbf {\bibinfo {volume} {3}},\ \bibinfo {pages}
  {3987} (\bibinfo {year} {2009})}\BibitemShut {NoStop}%
\bibitem [{\citenamefont {Vignolini}\ \emph {et~al.}(2012)\citenamefont
  {Vignolini}, \citenamefont {Yufa}, \citenamefont {Cunha}, \citenamefont
  {Guldin}, \citenamefont {Rushkin}, \citenamefont {Stefik}, \citenamefont
  {Hur}, \citenamefont {Wiesner}, \citenamefont {Baumberg},\ and\ \citenamefont
  {Steiner}}]{Vignolini2012}%
  \BibitemOpen
  \bibfield  {author} {\bibinfo {author} {\bibfnamefont {S.}~\bibnamefont
  {Vignolini}}, \bibinfo {author} {\bibfnamefont {N.~A.}\ \bibnamefont {Yufa}},
  \bibinfo {author} {\bibfnamefont {P.~S.}\ \bibnamefont {Cunha}}, \bibinfo
  {author} {\bibfnamefont {S.}~\bibnamefont {Guldin}}, \bibinfo {author}
  {\bibfnamefont {I.}~\bibnamefont {Rushkin}}, \bibinfo {author} {\bibfnamefont
  {M.}~\bibnamefont {Stefik}}, \bibinfo {author} {\bibfnamefont
  {K.}~\bibnamefont {Hur}}, \bibinfo {author} {\bibfnamefont {U.}~\bibnamefont
  {Wiesner}}, \bibinfo {author} {\bibfnamefont {J.~J.}\ \bibnamefont
  {Baumberg}},\ and\ \bibinfo {author} {\bibfnamefont {U.}~\bibnamefont
  {Steiner}},\ }\bibfield  {title} {\bibinfo {title} {A 3d optical metamaterial
  made by self-assembly},\ }\href
  {https://doi.org/https://doi.org/10.1002/adma.201103610} {\bibfield
  {journal} {\bibinfo  {journal} {Advanced Materials}\ }\textbf {\bibinfo
  {volume} {24}},\ \bibinfo {pages} {OP23} (\bibinfo {year}
  {2012})}\BibitemShut {NoStop}%
\bibitem [{\citenamefont {Hyde}\ \emph {et~al.}(1997)\citenamefont {Hyde},
  \citenamefont {Ninham}, \citenamefont {Andersson}, \citenamefont {Larsson},
  \citenamefont {Landh}, \citenamefont {Blum},\ and\ \citenamefont
  {Lidin}}]{Hyde1997_ch4}%
  \BibitemOpen
  \bibfield  {author} {\bibinfo {author} {\bibfnamefont {S.}~\bibnamefont
  {Hyde}}, \bibinfo {author} {\bibfnamefont {B.~W.}\ \bibnamefont {Ninham}},
  \bibinfo {author} {\bibfnamefont {S.}~\bibnamefont {Andersson}}, \bibinfo
  {author} {\bibfnamefont {K.}~\bibnamefont {Larsson}}, \bibinfo {author}
  {\bibfnamefont {T.}~\bibnamefont {Landh}}, \bibinfo {author} {\bibfnamefont
  {Z.}~\bibnamefont {Blum}},\ and\ \bibinfo {author} {\bibfnamefont
  {S.}~\bibnamefont {Lidin}},\ }\bibfield  {title} {\bibinfo {title} {Chapter 4
  - beyond flatland: The geometric forms due to self-assembly},\ }in\ \href
  {https://doi.org/https://doi.org/10.1016/B978-044481538-5/50005-8} {\emph
  {\bibinfo {booktitle} {The Language of Shape}}},\ \bibinfo {editor} {edited
  by\ \bibinfo {editor} {\bibfnamefont {S.}~\bibnamefont {Hyde}}, \bibinfo
  {editor} {\bibfnamefont {B.~W.}\ \bibnamefont {Ninham}}, \bibinfo {editor}
  {\bibfnamefont {S.}~\bibnamefont {Andersson}}, \bibinfo {editor}
  {\bibfnamefont {K.}~\bibnamefont {Larsson}}, \bibinfo {editor} {\bibfnamefont
  {T.}~\bibnamefont {Landh}}, \bibinfo {editor} {\bibfnamefont
  {Z.}~\bibnamefont {Blum}},\ and\ \bibinfo {editor} {\bibfnamefont
  {S.}~\bibnamefont {Lidin}}}\ (\bibinfo  {publisher} {Elsevier Science B.V.},\
  \bibinfo {address} {Amsterdam},\ \bibinfo {year} {1997})\ pp.\ \bibinfo
  {pages} {141--197}\BibitemShut {NoStop}%
\bibitem [{\citenamefont {Reddy}\ \emph {et~al.}(2021)\citenamefont {Reddy},
  \citenamefont {Feng}, \citenamefont {Thomas},\ and\ \citenamefont
  {Grason}}]{Reddy2021}%
  \BibitemOpen
  \bibfield  {author} {\bibinfo {author} {\bibfnamefont {A.}~\bibnamefont
  {Reddy}}, \bibinfo {author} {\bibfnamefont {X.}~\bibnamefont {Feng}},
  \bibinfo {author} {\bibfnamefont {E.~L.}\ \bibnamefont {Thomas}},\ and\
  \bibinfo {author} {\bibfnamefont {G.~M.}\ \bibnamefont {Grason}},\ }\bibfield
   {title} {\bibinfo {title} {Block copolymers beneath the surface: Measuring
  and modeling complex morphology at the subdomain scale},\ }\href
  {https://doi.org/10.1021/acs.macromol.1c00958} {\bibfield  {journal}
  {\bibinfo  {journal} {Macromolecules}\ }\textbf {\bibinfo {volume} {54}},\
  \bibinfo {pages} {9223} (\bibinfo {year} {2021})}\BibitemShut {NoStop}%
\bibitem [{\citenamefont {Grason}\ and\ \citenamefont
  {Thomas}(2023)}]{GrasonThomas2023}%
  \BibitemOpen
  \bibfield  {author} {\bibinfo {author} {\bibfnamefont {G.~M.}\ \bibnamefont
  {Grason}}\ and\ \bibinfo {author} {\bibfnamefont {E.~L.}\ \bibnamefont
  {Thomas}},\ }\bibfield  {title} {\bibinfo {title} {How does your gyroid grow?
  a mesoatomic perspective on supramolecular, soft matter network crystals},\
  }\href {https://doi.org/10.1103/PhysRevMaterials.7.045603} {\bibfield
  {journal} {\bibinfo  {journal} {Phys. Rev. Mater.}\ }\textbf {\bibinfo
  {volume} {7}},\ \bibinfo {pages} {045603} (\bibinfo {year}
  {2023})}\BibitemShut {NoStop}%
\bibitem [{\citenamefont {Feng}\ \emph {et~al.}(2019)\citenamefont {Feng},
  \citenamefont {Burke}, \citenamefont {Zhou}, \citenamefont {Guo},
  \citenamefont {Yang}, \citenamefont {Reddy}, \citenamefont {Prasad},
  \citenamefont {Ho}, \citenamefont {Avgeropoulos}, \citenamefont {Grason},\
  and\ \citenamefont {Thomas}}]{Feng2019}%
  \BibitemOpen
  \bibfield  {author} {\bibinfo {author} {\bibfnamefont {X.}~\bibnamefont
  {Feng}}, \bibinfo {author} {\bibfnamefont {C.~J.}\ \bibnamefont {Burke}},
  \bibinfo {author} {\bibfnamefont {M.}~\bibnamefont {Zhou}}, \bibinfo {author}
  {\bibfnamefont {H.}~\bibnamefont {Guo}}, \bibinfo {author} {\bibfnamefont
  {K.}~\bibnamefont {Yang}}, \bibinfo {author} {\bibfnamefont {A.}~\bibnamefont
  {Reddy}}, \bibinfo {author} {\bibfnamefont {I.}~\bibnamefont {Prasad}},
  \bibinfo {author} {\bibfnamefont {R.-M.}\ \bibnamefont {Ho}}, \bibinfo
  {author} {\bibfnamefont {A.}~\bibnamefont {Avgeropoulos}}, \bibinfo {author}
  {\bibfnamefont {G.~M.}\ \bibnamefont {Grason}},\ and\ \bibinfo {author}
  {\bibfnamefont {E.~L.}\ \bibnamefont {Thomas}},\ }\bibfield  {title}
  {\bibinfo {title} {Seeing mesoatomic distortions in soft-matter crystals of a
  double-gyroid block copolymer},\ }\href
  {https://doi.org/10.1038/s41586-019-1706-1} {\bibfield  {journal} {\bibinfo
  {journal} {Nature}\ }\textbf {\bibinfo {volume} {575}},\ \bibinfo {pages}
  {175–179} (\bibinfo {year} {2019})}\BibitemShut {NoStop}%
\bibitem [{\citenamefont {Feng}\ \emph {et~al.}(2023)\citenamefont {Feng},
  \citenamefont {Dimitriyev},\ and\ \citenamefont {Thomas}}]{Feng2023}%
  \BibitemOpen
  \bibfield  {author} {\bibinfo {author} {\bibfnamefont {X.}~\bibnamefont
  {Feng}}, \bibinfo {author} {\bibfnamefont {M.~S.}\ \bibnamefont
  {Dimitriyev}},\ and\ \bibinfo {author} {\bibfnamefont {E.~L.}\ \bibnamefont
  {Thomas}},\ }\bibfield  {title} {\bibinfo {title} {Soft, malleable double
  diamond twin},\ }\href {https://doi.org/10.1073/pnas.2213441120} {\bibfield
  {journal} {\bibinfo  {journal} {Proceedings of the National Academy of
  Sciences}\ }\textbf {\bibinfo {volume} {120}},\ \bibinfo {pages}
  {e2213441120} (\bibinfo {year} {2023})}\BibitemShut {NoStop}%
\bibitem [{\citenamefont {Durian}(1995)}]{Durian1995}%
  \BibitemOpen
  \bibfield  {author} {\bibinfo {author} {\bibfnamefont {D.~J.}\ \bibnamefont
  {Durian}},\ }\bibfield  {title} {\bibinfo {title} {Foam mechanics at the
  bubble scale},\ }\href {https://doi.org/10.1103/PhysRevLett.75.4780}
  {\bibfield  {journal} {\bibinfo  {journal} {Phys. Rev. Lett.}\ }\textbf
  {\bibinfo {volume} {75}},\ \bibinfo {pages} {4780} (\bibinfo {year}
  {1995})}\BibitemShut {NoStop}%
\bibitem [{\citenamefont {Langer}\ and\ \citenamefont
  {Liu}(1997)}]{Langer1997}%
  \BibitemOpen
  \bibfield  {author} {\bibinfo {author} {\bibfnamefont {S.~A.}\ \bibnamefont
  {Langer}}\ and\ \bibinfo {author} {\bibfnamefont {A.~J.}\ \bibnamefont
  {Liu}},\ }\bibfield  {title} {\bibinfo {title} {Effect of random packing on
  stress relaxation in foam},\ }\href {https://doi.org/10.1021/jp971265b}
  {\bibfield  {journal} {\bibinfo  {journal} {The Journal of Physical Chemistry
  B}\ }\textbf {\bibinfo {volume} {101}},\ \bibinfo {pages} {8667} (\bibinfo
  {year} {1997})}\BibitemShut {NoStop}%
\bibitem [{\citenamefont {Head}\ \emph {et~al.}(2003)\citenamefont {Head},
  \citenamefont {Levine},\ and\ \citenamefont {MacKintosh}}]{Head2003}%
  \BibitemOpen
  \bibfield  {author} {\bibinfo {author} {\bibfnamefont {D.~A.}\ \bibnamefont
  {Head}}, \bibinfo {author} {\bibfnamefont {A.~J.}\ \bibnamefont {Levine}},\
  and\ \bibinfo {author} {\bibfnamefont {F.~C.}\ \bibnamefont {MacKintosh}},\
  }\bibfield  {title} {\bibinfo {title} {Deformation of cross-linked
  semiflexible polymer networks},\ }\href
  {https://doi.org/10.1103/PhysRevLett.91.108102} {\bibfield  {journal}
  {\bibinfo  {journal} {Phys. Rev. Lett.}\ }\textbf {\bibinfo {volume} {91}},\
  \bibinfo {pages} {108102} (\bibinfo {year} {2003})}\BibitemShut {NoStop}%
\bibitem [{\citenamefont {DiDonna}\ and\ \citenamefont
  {Lubensky}(2005)}]{DiDonna2005}%
  \BibitemOpen
  \bibfield  {author} {\bibinfo {author} {\bibfnamefont {B.~A.}\ \bibnamefont
  {DiDonna}}\ and\ \bibinfo {author} {\bibfnamefont {T.~C.}\ \bibnamefont
  {Lubensky}},\ }\bibfield  {title} {\bibinfo {title} {Nonaffine correlations
  in random elastic media},\ }\href
  {https://doi.org/10.1103/PhysRevE.72.066619} {\bibfield  {journal} {\bibinfo
  {journal} {Phys. Rev. E}\ }\textbf {\bibinfo {volume} {72}},\ \bibinfo
  {pages} {066619} (\bibinfo {year} {2005})}\BibitemShut {NoStop}%
\bibitem [{\citenamefont {Conti}\ and\ \citenamefont
  {MacKintosh}(2009)}]{Conti2009}%
  \BibitemOpen
  \bibfield  {author} {\bibinfo {author} {\bibfnamefont {E.}~\bibnamefont
  {Conti}}\ and\ \bibinfo {author} {\bibfnamefont {F.~C.}\ \bibnamefont
  {MacKintosh}},\ }\bibfield  {title} {\bibinfo {title} {Cross-linked networks
  of stiff filaments exhibit negative normal stress},\ }\href
  {https://doi.org/10.1103/PhysRevLett.102.088102} {\bibfield  {journal}
  {\bibinfo  {journal} {Phys. Rev. Lett.}\ }\textbf {\bibinfo {volume} {102}},\
  \bibinfo {pages} {088102} (\bibinfo {year} {2009})}\BibitemShut {NoStop}%
\bibitem [{\citenamefont {Broedersz}\ \emph {et~al.}(2011)\citenamefont
  {Broedersz}, \citenamefont {Mao}, \citenamefont {Lubensky},\ and\
  \citenamefont {MacKintosh}}]{Broedersz2011}%
  \BibitemOpen
  \bibfield  {author} {\bibinfo {author} {\bibfnamefont {C.~P.}\ \bibnamefont
  {Broedersz}}, \bibinfo {author} {\bibfnamefont {X.}~\bibnamefont {Mao}},
  \bibinfo {author} {\bibfnamefont {T.~C.}\ \bibnamefont {Lubensky}},\ and\
  \bibinfo {author} {\bibfnamefont {F.~C.}\ \bibnamefont {MacKintosh}},\
  }\bibfield  {title} {\bibinfo {title} {Criticality and isostaticity in fibre
  networks},\ }\href {https://doi.org/10.1038/nphys2127} {\bibfield  {journal}
  {\bibinfo  {journal} {Nature Physics}\ }\textbf {\bibinfo {volume} {7}},\
  \bibinfo {pages} {983} (\bibinfo {year} {2011})}\BibitemShut {NoStop}%
\bibitem [{\citenamefont {Basu}\ \emph {et~al.}(2011)\citenamefont {Basu},
  \citenamefont {Wen}, \citenamefont {Mao}, \citenamefont {Lubensky},
  \citenamefont {Janmey},\ and\ \citenamefont {Yodh}}]{Basu2011}%
  \BibitemOpen
  \bibfield  {author} {\bibinfo {author} {\bibfnamefont {A.}~\bibnamefont
  {Basu}}, \bibinfo {author} {\bibfnamefont {Q.}~\bibnamefont {Wen}}, \bibinfo
  {author} {\bibfnamefont {X.}~\bibnamefont {Mao}}, \bibinfo {author}
  {\bibfnamefont {T.~C.}\ \bibnamefont {Lubensky}}, \bibinfo {author}
  {\bibfnamefont {P.~A.}\ \bibnamefont {Janmey}},\ and\ \bibinfo {author}
  {\bibfnamefont {A.~G.}\ \bibnamefont {Yodh}},\ }\bibfield  {title} {\bibinfo
  {title} {Nonaffine displacements in flexible polymer networks},\ }\href
  {https://doi.org/10.1021/ma1026803} {\bibfield  {journal} {\bibinfo
  {journal} {Macromolecules}\ }\textbf {\bibinfo {volume} {44}},\ \bibinfo
  {pages} {1671} (\bibinfo {year} {2011})}\BibitemShut {NoStop}%
\bibitem [{\citenamefont {Huisman}\ and\ \citenamefont
  {Lubensky}(2011)}]{Huisman2011}%
  \BibitemOpen
  \bibfield  {author} {\bibinfo {author} {\bibfnamefont {E.~M.}\ \bibnamefont
  {Huisman}}\ and\ \bibinfo {author} {\bibfnamefont {T.~C.}\ \bibnamefont
  {Lubensky}},\ }\bibfield  {title} {\bibinfo {title} {Internal stresses,
  normal modes, and nonaffinity in three-dimensional biopolymer networks},\
  }\href {https://doi.org/10.1103/PhysRevLett.106.088301} {\bibfield  {journal}
  {\bibinfo  {journal} {Phys. Rev. Lett.}\ }\textbf {\bibinfo {volume} {106}},\
  \bibinfo {pages} {088301} (\bibinfo {year} {2011})}\BibitemShut {NoStop}%
\bibitem [{\citenamefont {Broedersz}\ and\ \citenamefont
  {MacKintosh}(2014)}]{Broedersz2014}%
  \BibitemOpen
  \bibfield  {author} {\bibinfo {author} {\bibfnamefont {C.~P.}\ \bibnamefont
  {Broedersz}}\ and\ \bibinfo {author} {\bibfnamefont {F.~C.}\ \bibnamefont
  {MacKintosh}},\ }\bibfield  {title} {\bibinfo {title} {Modeling semiflexible
  polymer networks},\ }\href {https://doi.org/10.1103/RevModPhys.86.995}
  {\bibfield  {journal} {\bibinfo  {journal} {Rev. Mod. Phys.}\ }\textbf
  {\bibinfo {volume} {86}},\ \bibinfo {pages} {995} (\bibinfo {year}
  {2014})}\BibitemShut {NoStop}%
\bibitem [{\citenamefont {Feng}\ \emph {et~al.}(2016)\citenamefont {Feng},
  \citenamefont {Levine}, \citenamefont {Mao},\ and\ \citenamefont
  {Sander}}]{Feng2016}%
  \BibitemOpen
  \bibfield  {author} {\bibinfo {author} {\bibfnamefont {J.}~\bibnamefont
  {Feng}}, \bibinfo {author} {\bibfnamefont {H.}~\bibnamefont {Levine}},
  \bibinfo {author} {\bibfnamefont {X.}~\bibnamefont {Mao}},\ and\ \bibinfo
  {author} {\bibfnamefont {L.~M.}\ \bibnamefont {Sander}},\ }\bibfield  {title}
  {\bibinfo {title} {Nonlinear elasticity of disordered fiber networks},\
  }\href {https://doi.org/10.1039/C5SM01856K} {\bibfield  {journal} {\bibinfo
  {journal} {Soft Matter}\ }\textbf {\bibinfo {volume} {12}},\ \bibinfo {pages}
  {1419} (\bibinfo {year} {2016})}\BibitemShut {NoStop}%
\bibitem [{\citenamefont {Prakashchand}\ \emph {et~al.}(2020)\citenamefont
  {Prakashchand}, \citenamefont {Ahalawat}, \citenamefont {Bandyopadhyay},
  \citenamefont {Sengupta},\ and\ \citenamefont {Mondal}}]{Prakaschchand2020}%
  \BibitemOpen
  \bibfield  {author} {\bibinfo {author} {\bibfnamefont {D.~D.}\ \bibnamefont
  {Prakashchand}}, \bibinfo {author} {\bibfnamefont {N.}~\bibnamefont
  {Ahalawat}}, \bibinfo {author} {\bibfnamefont {S.}~\bibnamefont
  {Bandyopadhyay}}, \bibinfo {author} {\bibfnamefont {S.}~\bibnamefont
  {Sengupta}},\ and\ \bibinfo {author} {\bibfnamefont {J.}~\bibnamefont
  {Mondal}},\ }\bibfield  {title} {\bibinfo {title} {Nonaffine displacements
  encode collective conformational fluctuations in proteins},\ }\href
  {https://doi.org/10.1021/acs.jctc.9b01100} {\bibfield  {journal} {\bibinfo
  {journal} {Journal of Chemical Theory and Computation}\ }\textbf {\bibinfo
  {volume} {16}},\ \bibinfo {pages} {2508} (\bibinfo {year}
  {2020})}\BibitemShut {NoStop}%
\bibitem [{\citenamefont {Pensalfini}\ \emph {et~al.}(2023)\citenamefont
  {Pensalfini}, \citenamefont {Golde}, \citenamefont {Trepat},\ and\
  \citenamefont {Arroyo}}]{Pensalfini2023}%
  \BibitemOpen
  \bibfield  {author} {\bibinfo {author} {\bibfnamefont {M.}~\bibnamefont
  {Pensalfini}}, \bibinfo {author} {\bibfnamefont {T.}~\bibnamefont {Golde}},
  \bibinfo {author} {\bibfnamefont {X.}~\bibnamefont {Trepat}},\ and\ \bibinfo
  {author} {\bibfnamefont {M.}~\bibnamefont {Arroyo}},\ }\bibfield  {title}
  {\bibinfo {title} {Nonaffine mechanics of entangled networks inspired by
  intermediate filaments},\ }\href
  {https://doi.org/10.1103/PhysRevLett.131.058101} {\bibfield  {journal}
  {\bibinfo  {journal} {Phys. Rev. Lett.}\ }\textbf {\bibinfo {volume} {131}},\
  \bibinfo {pages} {058101} (\bibinfo {year} {2023})}\BibitemShut {NoStop}%
\bibitem [{\citenamefont {Jari\ifmmode~\acute{c}\else \'{c}\fi{}}\ and\
  \citenamefont {Mohanty}(1988)}]{Jaric1988}%
  \BibitemOpen
  \bibfield  {author} {\bibinfo {author} {\bibfnamefont {M.~V.}\ \bibnamefont
  {Jari\ifmmode~\acute{c}\else \'{c}\fi{}}}\ and\ \bibinfo {author}
  {\bibfnamefont {U.}~\bibnamefont {Mohanty}},\ }\bibfield  {title} {\bibinfo
  {title} {Density-functional theory of elastic moduli: Hard-sphere and
  lennard-jones crystals},\ }\href {https://doi.org/10.1103/PhysRevB.37.4441}
  {\bibfield  {journal} {\bibinfo  {journal} {Phys. Rev. B}\ }\textbf {\bibinfo
  {volume} {37}},\ \bibinfo {pages} {4441} (\bibinfo {year}
  {1988})}\BibitemShut {NoStop}%
\bibitem [{\citenamefont {Ganguly}\ \emph {et~al.}(2013)\citenamefont
  {Ganguly}, \citenamefont {Sengupta}, \citenamefont {Sollich},\ and\
  \citenamefont {Rao}}]{Ganguly2013}%
  \BibitemOpen
  \bibfield  {author} {\bibinfo {author} {\bibfnamefont {S.}~\bibnamefont
  {Ganguly}}, \bibinfo {author} {\bibfnamefont {S.}~\bibnamefont {Sengupta}},
  \bibinfo {author} {\bibfnamefont {P.}~\bibnamefont {Sollich}},\ and\ \bibinfo
  {author} {\bibfnamefont {M.}~\bibnamefont {Rao}},\ }\bibfield  {title}
  {\bibinfo {title} {Nonaffine displacements in crystalline solids in the
  harmonic limit},\ }\href {https://doi.org/10.1103/PhysRevE.87.042801}
  {\bibfield  {journal} {\bibinfo  {journal} {Phys. Rev. E}\ }\textbf {\bibinfo
  {volume} {87}},\ \bibinfo {pages} {042801} (\bibinfo {year}
  {2013})}\BibitemShut {NoStop}%
\bibitem [{\citenamefont {Das}\ \emph {et~al.}(2015)\citenamefont {Das},
  \citenamefont {Ganguly}, \citenamefont {Sengupta},\ and\ \citenamefont
  {Rao}}]{Das2015}%
  \BibitemOpen
  \bibfield  {author} {\bibinfo {author} {\bibfnamefont {T.}~\bibnamefont
  {Das}}, \bibinfo {author} {\bibfnamefont {S.}~\bibnamefont {Ganguly}},
  \bibinfo {author} {\bibfnamefont {S.}~\bibnamefont {Sengupta}},\ and\
  \bibinfo {author} {\bibfnamefont {M.}~\bibnamefont {Rao}},\ }\bibfield
  {title} {\bibinfo {title} {Pre-yield non-affine fluctuations and a hidden
  critical point in strained crystals},\ }\href
  {https://doi.org/10.1038/srep10644} {\bibfield  {journal} {\bibinfo
  {journal} {Scientific Reports}\ }\textbf {\bibinfo {volume} {5}},\ \bibinfo
  {pages} {10644} (\bibinfo {year} {2015})}\BibitemShut {NoStop}%
\bibitem [{\citenamefont {Fruchart}\ \emph {et~al.}(2018)\citenamefont
  {Fruchart}, \citenamefont {Jeon}, \citenamefont {Hur}, \citenamefont
  {Cheianov}, \citenamefont {Wiesner},\ and\ \citenamefont
  {Vitelli}}]{Fruchart2018}%
  \BibitemOpen
  \bibfield  {author} {\bibinfo {author} {\bibfnamefont {M.}~\bibnamefont
  {Fruchart}}, \bibinfo {author} {\bibfnamefont {S.-Y.}\ \bibnamefont {Jeon}},
  \bibinfo {author} {\bibfnamefont {K.}~\bibnamefont {Hur}}, \bibinfo {author}
  {\bibfnamefont {V.}~\bibnamefont {Cheianov}}, \bibinfo {author}
  {\bibfnamefont {U.}~\bibnamefont {Wiesner}},\ and\ \bibinfo {author}
  {\bibfnamefont {V.}~\bibnamefont {Vitelli}},\ }\bibfield  {title} {\bibinfo
  {title} {Soft self-assembly of weyl materials for light and sound},\ }\href
  {https://doi.org/10.1073/pnas.1720828115} {\bibfield  {journal} {\bibinfo
  {journal} {Proceedings of the National Academy of Sciences}\ }\textbf
  {\bibinfo {volume} {115}},\ \bibinfo {pages} {E3655} (\bibinfo {year}
  {2018})}\BibitemShut {NoStop}%
\bibitem [{\citenamefont {Jo}\ \emph {et~al.}(2021)\citenamefont {Jo},
  \citenamefont {Park}, \citenamefont {Jun}, \citenamefont {Kim}, \citenamefont
  {Jung}, \citenamefont {Park}, \citenamefont {Lee}, \citenamefont {Lee},\ and\
  \citenamefont {Ryu}}]{Jo2021}%
  \BibitemOpen
  \bibfield  {author} {\bibinfo {author} {\bibfnamefont {S.}~\bibnamefont
  {Jo}}, \bibinfo {author} {\bibfnamefont {H.}~\bibnamefont {Park}}, \bibinfo
  {author} {\bibfnamefont {T.}~\bibnamefont {Jun}}, \bibinfo {author}
  {\bibfnamefont {K.}~\bibnamefont {Kim}}, \bibinfo {author} {\bibfnamefont
  {H.}~\bibnamefont {Jung}}, \bibinfo {author} {\bibfnamefont {S.}~\bibnamefont
  {Park}}, \bibinfo {author} {\bibfnamefont {B.}~\bibnamefont {Lee}}, \bibinfo
  {author} {\bibfnamefont {S.}~\bibnamefont {Lee}},\ and\ \bibinfo {author}
  {\bibfnamefont {D.~Y.}\ \bibnamefont {Ryu}},\ }\bibfield  {title} {\bibinfo
  {title} {Symmetry-breaking in double gyroid block copolymer films by
  non-affine distortion},\ }\href
  {https://doi.org/https://doi.org/10.1016/j.apmt.2021.101006} {\bibfield
  {journal} {\bibinfo  {journal} {Applied Materials Today}\ }\textbf {\bibinfo
  {volume} {23}},\ \bibinfo {pages} {101006} (\bibinfo {year}
  {2021})}\BibitemShut {NoStop}%
\bibitem [{\citenamefont {Park}\ \emph {et~al.}(2022)\citenamefont {Park},
  \citenamefont {Jo}, \citenamefont {Kang}, \citenamefont {Hur}, \citenamefont
  {Oh}, \citenamefont {Ryu},\ and\ \citenamefont {Lee}}]{Park2022}%
  \BibitemOpen
  \bibfield  {author} {\bibinfo {author} {\bibfnamefont {H.}~\bibnamefont
  {Park}}, \bibinfo {author} {\bibfnamefont {S.}~\bibnamefont {Jo}}, \bibinfo
  {author} {\bibfnamefont {B.}~\bibnamefont {Kang}}, \bibinfo {author}
  {\bibfnamefont {K.}~\bibnamefont {Hur}}, \bibinfo {author} {\bibfnamefont
  {S.~S.}\ \bibnamefont {Oh}}, \bibinfo {author} {\bibfnamefont {D.~Y.}\
  \bibnamefont {Ryu}},\ and\ \bibinfo {author} {\bibfnamefont {S.}~\bibnamefont
  {Lee}},\ }\bibfield  {title} {\bibinfo {title} {Block copolymer gyroids for
  nanophotonics: significance of lattice transformations},\ }\href
  {https://doi.org/doi:10.1515/nanoph-2021-0644} {\bibfield  {journal}
  {\bibinfo  {journal} {Nanophotonics}\ }\textbf {\bibinfo {volume} {11}},\
  \bibinfo {pages} {2583} (\bibinfo {year} {2022})}\BibitemShut {NoStop}%
\bibitem [{Note1()}]{Note1}%
  \BibitemOpen
  \bibinfo {note} {We opt to describe nodes as polyhedra, where each strut
  emerging from the node is represented by a face. DG nodes are described as
  trihedral; DD nodes are tetrahedral; DP nodes are hexahedral (described by
  octahedral symmetry group $O_h$).}\BibitemShut {Stop}%
\bibitem [{\citenamefont {Wells}(1977)}]{Wells1977}%
  \BibitemOpen
  \bibfield  {author} {\bibinfo {author} {\bibfnamefont {A.~F.}\ \bibnamefont
  {Wells}},\ }\href@noop {} {\emph {\bibinfo {title} {Three Dimensional Nets
  and Polyhedra}}}\ (\bibinfo  {publisher} {Wiley},\ \bibinfo {address} {New
  York},\ \bibinfo {year} {1977})\BibitemShut {NoStop}%
\bibitem [{\citenamefont {Anderson}\ \emph {et~al.}(1988)\citenamefont
  {Anderson}, \citenamefont {Gruner},\ and\ \citenamefont
  {Leibler}}]{Anderson1988}%
  \BibitemOpen
  \bibfield  {author} {\bibinfo {author} {\bibfnamefont {D.~M.}\ \bibnamefont
  {Anderson}}, \bibinfo {author} {\bibfnamefont {S.~M.}\ \bibnamefont
  {Gruner}},\ and\ \bibinfo {author} {\bibfnamefont {S.}~\bibnamefont
  {Leibler}},\ }\bibfield  {title} {\bibinfo {title} {{Geometrical aspects of
  the frustration in the cubic phases of lyotropic liquid crystals.}},\ }\href
  {https://doi.org/10.1073/pnas.85.15.5364} {\bibfield  {journal} {\bibinfo
  {journal} {Proceedings of the National Academy of Sciences}\ }\textbf
  {\bibinfo {volume} {85}},\ \bibinfo {pages} {5364} (\bibinfo {year}
  {1988})}\BibitemShut {NoStop}%
\bibitem [{\citenamefont {Matsen}\ and\ \citenamefont
  {Bates}(1996)}]{Matsen1996}%
  \BibitemOpen
  \bibfield  {author} {\bibinfo {author} {\bibfnamefont {M.~W.}\ \bibnamefont
  {Matsen}}\ and\ \bibinfo {author} {\bibfnamefont {F.~S.}\ \bibnamefont
  {Bates}},\ }\bibfield  {title} {\bibinfo {title} {{Origins of Complex
  Self-Assembly in Block Copolymers}},\ }\href
  {https://doi.org/10.1021/ma960744q} {\bibfield  {journal} {\bibinfo
  {journal} {Macromolecules}\ }\textbf {\bibinfo {volume} {29}},\ \bibinfo
  {pages} {7641} (\bibinfo {year} {1996})}\BibitemShut {NoStop}%
\bibitem [{\citenamefont {Alex}\ and\ \citenamefont
  {Grosse-Brauckmann}(2019)}]{Alex2019}%
  \BibitemOpen
  \bibfield  {author} {\bibinfo {author} {\bibfnamefont {J.}~\bibnamefont
  {Alex}}\ and\ \bibinfo {author} {\bibfnamefont {K.}~\bibnamefont
  {Grosse-Brauckmann}},\ }\href@noop {} {\bibinfo {title} {Periodic networks of
  fixed degree minimizing length}} (\bibinfo {year} {2019}),\ \Eprint
  {https://arxiv.org/abs/1911.01792} {arXiv:1911.01792 [math.CO]} \BibitemShut
  {NoStop}%
\bibitem [{\citenamefont {Alex}\ and\ \citenamefont
  {Grosse-Brauckmann}(2023)}]{Alex2023}%
  \BibitemOpen
  \bibfield  {author} {\bibinfo {author} {\bibfnamefont {J.}~\bibnamefont
  {Alex}}\ and\ \bibinfo {author} {\bibfnamefont {K.}~\bibnamefont
  {Grosse-Brauckmann}},\ }\bibfield  {title} {\bibinfo {title} {Periodic
  steiner networks minimizing length},\ }\bibfield  {journal} {\bibinfo
  {journal} {Discrete \& Computational Geometry}\ }\href
  {https://doi.org/10.1007/s00454-023-00576-z} {10.1007/s00454-023-00576-z}
  (\bibinfo {year} {2023})\BibitemShut {NoStop}%
\bibitem [{\citenamefont {Isenberg}(1992)}]{Isenberg1992}%
  \BibitemOpen
  \bibfield  {author} {\bibinfo {author} {\bibfnamefont {C.}~\bibnamefont
  {Isenberg}},\ }\href@noop {} {\emph {\bibinfo {title} {{The Science of Soap
  Films and Soap Bubbles}}}},\ Dover books explaining science\ (\bibinfo
  {publisher} {Dover Publications},\ \bibinfo {year} {1992})\BibitemShut
  {NoStop}%
\bibitem [{\citenamefont {Ivanov}\ and\ \citenamefont
  {Tuzhilin}(1994)}]{Ivanov1994MinimalNT}%
  \BibitemOpen
  \bibfield  {author} {\bibinfo {author} {\bibfnamefont {A.}~\bibnamefont
  {Ivanov}}\ and\ \bibinfo {author} {\bibfnamefont {A.~A.}\ \bibnamefont
  {Tuzhilin}},\ }\bibinfo {title} {{Minimal Networks: The Steiner Problem and
  Its Generalizations}}\ (\bibinfo  {publisher} {CRC Press},\ \bibinfo {year}
  {1994})\BibitemShut {NoStop}%
\bibitem [{\citenamefont {Wyart}\ \emph {et~al.}(2008)\citenamefont {Wyart},
  \citenamefont {Liang}, \citenamefont {Kabla},\ and\ \citenamefont
  {Mahadevan}}]{Wyart2008}%
  \BibitemOpen
  \bibfield  {author} {\bibinfo {author} {\bibfnamefont {M.}~\bibnamefont
  {Wyart}}, \bibinfo {author} {\bibfnamefont {H.}~\bibnamefont {Liang}},
  \bibinfo {author} {\bibfnamefont {A.}~\bibnamefont {Kabla}},\ and\ \bibinfo
  {author} {\bibfnamefont {L.}~\bibnamefont {Mahadevan}},\ }\bibfield  {title}
  {\bibinfo {title} {Elasticity of floppy and stiff random networks},\ }\href
  {https://doi.org/10.1103/PhysRevLett.101.215501} {\bibfield  {journal}
  {\bibinfo  {journal} {Phys. Rev. Lett.}\ }\textbf {\bibinfo {volume} {101}},\
  \bibinfo {pages} {215501} (\bibinfo {year} {2008})}\BibitemShut {NoStop}%
\bibitem [{\citenamefont {Carrillo}\ \emph {et~al.}(2013)\citenamefont
  {Carrillo}, \citenamefont {MacKintosh},\ and\ \citenamefont
  {Dobrynin}}]{Carrillo2013}%
  \BibitemOpen
  \bibfield  {author} {\bibinfo {author} {\bibfnamefont {J.-M.~Y.}\
  \bibnamefont {Carrillo}}, \bibinfo {author} {\bibfnamefont {F.~C.}\
  \bibnamefont {MacKintosh}},\ and\ \bibinfo {author} {\bibfnamefont {A.~V.}\
  \bibnamefont {Dobrynin}},\ }\bibfield  {title} {\bibinfo {title} {Nonlinear
  elasticity: From single chain to networks and gels},\ }\href
  {https://doi.org/10.1021/ma400478f} {\bibfield  {journal} {\bibinfo
  {journal} {Macromolecules}\ }\textbf {\bibinfo {volume} {46}},\ \bibinfo
  {pages} {3679} (\bibinfo {year} {2013})}\BibitemShut {NoStop}%
\bibitem [{\citenamefont {Gardel}\ \emph {et~al.}(2004)\citenamefont {Gardel},
  \citenamefont {Shin}, \citenamefont {MacKintosh}, \citenamefont {Mahadevan},
  \citenamefont {Matsudaira},\ and\ \citenamefont {Weitz}}]{Gardel2004}%
  \BibitemOpen
  \bibfield  {author} {\bibinfo {author} {\bibfnamefont {M.~L.}\ \bibnamefont
  {Gardel}}, \bibinfo {author} {\bibfnamefont {J.~H.}\ \bibnamefont {Shin}},
  \bibinfo {author} {\bibfnamefont {F.~C.}\ \bibnamefont {MacKintosh}},
  \bibinfo {author} {\bibfnamefont {L.}~\bibnamefont {Mahadevan}}, \bibinfo
  {author} {\bibfnamefont {P.}~\bibnamefont {Matsudaira}},\ and\ \bibinfo
  {author} {\bibfnamefont {D.~A.}\ \bibnamefont {Weitz}},\ }\bibfield  {title}
  {\bibinfo {title} {Elastic behavior of cross-linked and bundled actin
  networks},\ }\href {https://doi.org/10.1126/science.1095087} {\bibfield
  {journal} {\bibinfo  {journal} {Science}\ }\textbf {\bibinfo {volume}
  {304}},\ \bibinfo {pages} {1301} (\bibinfo {year} {2004})},\ \Eprint
  {https://arxiv.org/abs/https://www.science.org/doi/pdf/10.1126/science.1095087}
  {https://www.science.org/doi/pdf/10.1126/science.1095087} \BibitemShut
  {NoStop}%
\bibitem [{\citenamefont {Storm}\ \emph {et~al.}(2005)\citenamefont {Storm},
  \citenamefont {Pastore}, \citenamefont {MacKintosh}, \citenamefont
  {Lubensky},\ and\ \citenamefont {Janmey}}]{Storm2005}%
  \BibitemOpen
  \bibfield  {author} {\bibinfo {author} {\bibfnamefont {C.}~\bibnamefont
  {Storm}}, \bibinfo {author} {\bibfnamefont {J.~J.}\ \bibnamefont {Pastore}},
  \bibinfo {author} {\bibfnamefont {F.~C.}\ \bibnamefont {MacKintosh}},
  \bibinfo {author} {\bibfnamefont {T.~C.}\ \bibnamefont {Lubensky}},\ and\
  \bibinfo {author} {\bibfnamefont {P.~A.}\ \bibnamefont {Janmey}},\ }\bibfield
   {title} {\bibinfo {title} {Nonlinear elasticity in biological gels},\ }\href
  {https://doi.org/10.1038/nature03521} {\bibfield  {journal} {\bibinfo
  {journal} {Nature}\ }\textbf {\bibinfo {volume} {435}},\ \bibinfo {pages}
  {191} (\bibinfo {year} {2005})}\BibitemShut {NoStop}%
\bibitem [{\citenamefont {Lieleg}\ \emph {et~al.}(2007)\citenamefont {Lieleg},
  \citenamefont {Claessens}, \citenamefont {Heussinger}, \citenamefont {Frey},\
  and\ \citenamefont {Bausch}}]{Lieleg2007}%
  \BibitemOpen
  \bibfield  {author} {\bibinfo {author} {\bibfnamefont {O.}~\bibnamefont
  {Lieleg}}, \bibinfo {author} {\bibfnamefont {M.~M. A.~E.}\ \bibnamefont
  {Claessens}}, \bibinfo {author} {\bibfnamefont {C.}~\bibnamefont
  {Heussinger}}, \bibinfo {author} {\bibfnamefont {E.}~\bibnamefont {Frey}},\
  and\ \bibinfo {author} {\bibfnamefont {A.~R.}\ \bibnamefont {Bausch}},\
  }\bibfield  {title} {\bibinfo {title} {Mechanics of bundled semiflexible
  polymer networks},\ }\href {https://doi.org/10.1103/PhysRevLett.99.088102}
  {\bibfield  {journal} {\bibinfo  {journal} {Phys. Rev. Lett.}\ }\textbf
  {\bibinfo {volume} {99}},\ \bibinfo {pages} {088102} (\bibinfo {year}
  {2007})}\BibitemShut {NoStop}%
\bibitem [{\citenamefont {Sharma}\ \emph {et~al.}(2016)\citenamefont {Sharma},
  \citenamefont {Licup}, \citenamefont {Jansen}, \citenamefont {Rens},
  \citenamefont {Sheinman}, \citenamefont {Koenderink},\ and\ \citenamefont
  {MacKintosh}}]{Sharma2016}%
  \BibitemOpen
  \bibfield  {author} {\bibinfo {author} {\bibfnamefont {A.}~\bibnamefont
  {Sharma}}, \bibinfo {author} {\bibfnamefont {A.~J.}\ \bibnamefont {Licup}},
  \bibinfo {author} {\bibfnamefont {K.~A.}\ \bibnamefont {Jansen}}, \bibinfo
  {author} {\bibfnamefont {R.}~\bibnamefont {Rens}}, \bibinfo {author}
  {\bibfnamefont {M.}~\bibnamefont {Sheinman}}, \bibinfo {author}
  {\bibfnamefont {G.~H.}\ \bibnamefont {Koenderink}},\ and\ \bibinfo {author}
  {\bibfnamefont {F.~C.}\ \bibnamefont {MacKintosh}},\ }\bibfield  {title}
  {\bibinfo {title} {Strain-controlled criticality governs the nonlinear
  mechanics of fibre networks},\ }\href {https://doi.org/10.1038/nphys3628}
  {\bibfield  {journal} {\bibinfo  {journal} {Nature Physics}\ }\textbf
  {\bibinfo {volume} {12}},\ \bibinfo {pages} {584} (\bibinfo {year}
  {2016})}\BibitemShut {NoStop}%
\bibitem [{\citenamefont {Marko}\ and\ \citenamefont
  {Siggia}(1995)}]{Marko1995}%
  \BibitemOpen
  \bibfield  {author} {\bibinfo {author} {\bibfnamefont {J.~F.}\ \bibnamefont
  {Marko}}\ and\ \bibinfo {author} {\bibfnamefont {E.~D.}\ \bibnamefont
  {Siggia}},\ }\bibfield  {title} {\bibinfo {title} {Stretching {DNA}},\
  }\href@noop {} {\bibfield  {journal} {\bibinfo  {journal} {Macromolecules}\
  }\textbf {\bibinfo {volume} {28}},\ \bibinfo {pages} {8759} (\bibinfo {year}
  {1995})}\BibitemShut {NoStop}%
\bibitem [{\citenamefont {Rubinstein}\ and\ \citenamefont
  {Colby}(2003)}]{Rubinstein2003}%
  \BibitemOpen
  \bibfield  {author} {\bibinfo {author} {\bibfnamefont {M.}~\bibnamefont
  {Rubinstein}}\ and\ \bibinfo {author} {\bibfnamefont {R.}~\bibnamefont
  {Colby}},\ }\href@noop {} {\emph {\bibinfo {title} {Polymer Physics}}}\
  (\bibinfo  {publisher} {Oxford University Press},\ \bibinfo {year}
  {2003})\BibitemShut {NoStop}%
\bibitem [{\citenamefont {Treloar}(1975)}]{Treloar1975}%
  \BibitemOpen
  \bibfield  {author} {\bibinfo {author} {\bibfnamefont {L.}~\bibnamefont
  {Treloar}},\ }\href@noop {} {\emph {\bibinfo {title} {The Physics of Rubber
  Elasticity}}},\ Oxford Classic Texts in the Physical Sciences\ (\bibinfo
  {publisher} {OUP Oxford},\ \bibinfo {year} {1975})\BibitemShut {NoStop}%
\bibitem [{\citenamefont {Flory}(1985)}]{Flory1985}%
  \BibitemOpen
  \bibfield  {author} {\bibinfo {author} {\bibfnamefont {P.~J.}\ \bibnamefont
  {Flory}},\ }\bibfield  {title} {\bibinfo {title} {Molecular theory of rubber
  elasticity},\ }\href {https://doi.org/10.1295/polymj.17.1} {\bibfield
  {journal} {\bibinfo  {journal} {Polymer Journal}\ }\textbf {\bibinfo {volume}
  {17}},\ \bibinfo {pages} {1} (\bibinfo {year} {1985})}\BibitemShut {NoStop}%
\bibitem [{\citenamefont {Alexander}(1998)}]{Alexander1998}%
  \BibitemOpen
  \bibfield  {author} {\bibinfo {author} {\bibfnamefont {S.}~\bibnamefont
  {Alexander}},\ }\bibfield  {title} {\bibinfo {title} {Amorphous solids: their
  structure, lattice dynamics and elasticity},\ }\href
  {https://doi.org/https://doi.org/10.1016/S0370-1573(97)00069-0} {\bibfield
  {journal} {\bibinfo  {journal} {Physics Reports}\ }\textbf {\bibinfo {volume}
  {296}},\ \bibinfo {pages} {65} (\bibinfo {year} {1998})}\BibitemShut
  {NoStop}%
\bibitem [{\citenamefont {Fischer}\ and\ \citenamefont
  {Koch}(1996)}]{Fischer1996}%
  \BibitemOpen
  \bibfield  {author} {\bibinfo {author} {\bibfnamefont {W.}~\bibnamefont
  {Fischer}}\ and\ \bibinfo {author} {\bibfnamefont {E.}~\bibnamefont {Koch}},\
  }\bibfield  {title} {\bibinfo {title} {Spanning minimal surfaces},\ }\href
  {https://doi.org/10.1098/rsta.1996.0094} {\bibfield  {journal} {\bibinfo
  {journal} {Philosophical Transactions of the Royal Society of London. Series
  A: Mathematical, Physical and Engineering Sciences}\ }\textbf {\bibinfo
  {volume} {354}},\ \bibinfo {pages} {2105} (\bibinfo {year}
  {1996})}\BibitemShut {NoStop}%
\bibitem [{\citenamefont {Landau}\ \emph {et~al.}(1986)\citenamefont {Landau},
  \citenamefont {Lifshitz}, \citenamefont {Kosevich},\ and\ \citenamefont
  {Pitaevskii}}]{LandauLifshitz1986}%
  \BibitemOpen
  \bibfield  {author} {\bibinfo {author} {\bibfnamefont {L.~D.}\ \bibnamefont
  {Landau}}, \bibinfo {author} {\bibfnamefont {E.~M.}\ \bibnamefont
  {Lifshitz}}, \bibinfo {author} {\bibfnamefont {A.~M.}\ \bibnamefont
  {Kosevich}},\ and\ \bibinfo {author} {\bibfnamefont {L.~P.}\ \bibnamefont
  {Pitaevskii}},\ }\href@noop {} {\emph {\bibinfo {title} {Theory of
  Elasticity}}},\ Course of theoretical physics\ (\bibinfo  {publisher}
  {Butterworth-Heinemann},\ \bibinfo {year} {1986})\BibitemShut {NoStop}%
\bibitem [{\citenamefont {Prasad}\ \emph {et~al.}(2018)\citenamefont {Prasad},
  \citenamefont {Jinnai}, \citenamefont {Ho}, \citenamefont {Thomas},\ and\
  \citenamefont {Grason}}]{Prasad2018}%
  \BibitemOpen
  \bibfield  {author} {\bibinfo {author} {\bibfnamefont {I.}~\bibnamefont
  {Prasad}}, \bibinfo {author} {\bibfnamefont {H.}~\bibnamefont {Jinnai}},
  \bibinfo {author} {\bibfnamefont {R.-M.}\ \bibnamefont {Ho}}, \bibinfo
  {author} {\bibfnamefont {E.~L.}\ \bibnamefont {Thomas}},\ and\ \bibinfo
  {author} {\bibfnamefont {G.~M.}\ \bibnamefont {Grason}},\ }\bibfield  {title}
  {\bibinfo {title} {Anatomy of triply-periodic network assemblies:
  characterizing skeletal and inter-domain surface geometry of block copolymer
  gyroids},\ }\href {https://doi.org/10.1039/C8SM00078F} {\bibfield  {journal}
  {\bibinfo  {journal} {Soft Matter}\ }\textbf {\bibinfo {volume} {14}},\
  \bibinfo {pages} {3612} (\bibinfo {year} {2018})}\BibitemShut {NoStop}%
\end{thebibliography}%


\begin{thebibliography}{15}%
\makeatletter
\providecommand \@ifxundefined [1]{%
 \@ifx{#1\undefined}
}%
\providecommand \@ifnum [1]{%
 \ifnum #1\expandafter \@firstoftwo
 \else \expandafter \@secondoftwo
 \fi
}%
\providecommand \@ifx [1]{%
 \ifx #1\expandafter \@firstoftwo
 \else \expandafter \@secondoftwo
 \fi
}%
\providecommand \natexlab [1]{#1}%
\providecommand \enquote  [1]{``#1''}%
\providecommand \bibnamefont  [1]{#1}%
\providecommand \bibfnamefont [1]{#1}%
\providecommand \citenamefont [1]{#1}%
\providecommand \href@noop [0]{\@secondoftwo}%
\providecommand \href [0]{\begingroup \@sanitize@url \@href}%
\providecommand \@href[1]{\@@startlink{#1}\@@href}%
\providecommand \@@href[1]{\endgroup#1\@@endlink}%
\providecommand \@sanitize@url [0]{\catcode `\\12\catcode `\$12\catcode
  `\&12\catcode `\#12\catcode `\^12\catcode `\_12\catcode `\%12\relax}%
\providecommand \@@startlink[1]{}%
\providecommand \@@endlink[0]{}%
\providecommand \url  [0]{\begingroup\@sanitize@url \@url }%
\providecommand \@url [1]{\endgroup\@href {#1}{\urlprefix }}%
\providecommand \urlprefix  [0]{URL }%
\providecommand \Eprint [0]{\href }%
\providecommand \doibase [0]{https://doi.org/}%
\providecommand \selectlanguage [0]{\@gobble}%
\providecommand \bibinfo  [0]{\@secondoftwo}%
\providecommand \bibfield  [0]{\@secondoftwo}%
\providecommand \translation [1]{[#1]}%
\providecommand \BibitemOpen [0]{}%
\providecommand \bibitemStop [0]{}%
\providecommand \bibitemNoStop [0]{.\EOS\space}%
\providecommand \EOS [0]{\spacefactor3000\relax}%
\providecommand \BibitemShut  [1]{\csname bibitem#1\endcsname}%
\let\auto@bib@innerbib\@empty
\bibitem [{\citenamefont {Ivanov}\ and\ \citenamefont
  {Tuzhilin}(1994)}]{Ivanov1994MinimalNT}%
  \BibitemOpen
  \bibfield  {author} {\bibinfo {author} {\bibfnamefont {A.}~\bibnamefont
  {Ivanov}}\ and\ \bibinfo {author} {\bibfnamefont {A.~A.}\ \bibnamefont
  {Tuzhilin}},\ }\bibinfo {title} {{Minimal Networks: The Steiner Problem and
  Its Generalizations}}\ (\bibinfo  {publisher} {CRC Press},\ \bibinfo {year}
  {1994})\BibitemShut {NoStop}%
\bibitem [{\citenamefont {Alex}\ and\ \citenamefont
  {Grosse-Brauckmann}(2019)}]{Alex2019}%
  \BibitemOpen
  \bibfield  {author} {\bibinfo {author} {\bibfnamefont {J.}~\bibnamefont
  {Alex}}\ and\ \bibinfo {author} {\bibfnamefont {K.}~\bibnamefont
  {Grosse-Brauckmann}},\ }\href@noop {} {\bibinfo {title} {Periodic networks of
  fixed degree minimizing length}} (\bibinfo {year} {2019}),\ \Eprint
  {https://arxiv.org/abs/1911.01792} {arXiv:1911.01792 [math.CO]} \BibitemShut
  {NoStop}%
\bibitem [{\citenamefont {Alex}\ and\ \citenamefont
  {Grosse-Brauckmann}(2023)}]{Alex2023}%
  \BibitemOpen
  \bibfield  {author} {\bibinfo {author} {\bibfnamefont {J.}~\bibnamefont
  {Alex}}\ and\ \bibinfo {author} {\bibfnamefont {K.}~\bibnamefont
  {Grosse-Brauckmann}},\ }\bibfield  {title} {\bibinfo {title} {Periodic
  steiner networks minimizing length},\ }\bibfield  {journal} {\bibinfo
  {journal} {Discrete \& Computational Geometry}\ }\href
  {https://doi.org/10.1007/s00454-023-00576-z} {10.1007/s00454-023-00576-z}
  (\bibinfo {year} {2023})\BibitemShut {NoStop}%
\bibitem [{\citenamefont {Marko}\ and\ \citenamefont
  {Siggia}(1995)}]{Marko1995}%
  \BibitemOpen
  \bibfield  {author} {\bibinfo {author} {\bibfnamefont {J.~F.}\ \bibnamefont
  {Marko}}\ and\ \bibinfo {author} {\bibfnamefont {E.~D.}\ \bibnamefont
  {Siggia}},\ }\bibfield  {title} {\bibinfo {title} {Stretching {DNA}},\
  }\href@noop {} {\bibfield  {journal} {\bibinfo  {journal} {Macromolecules}\
  }\textbf {\bibinfo {volume} {28}},\ \bibinfo {pages} {8759} (\bibinfo {year}
  {1995})}\BibitemShut {NoStop}%
\bibitem [{\citenamefont {Rubinstein}\ and\ \citenamefont
  {Colby}(2003)}]{Rubinstein2003}%
  \BibitemOpen
  \bibfield  {author} {\bibinfo {author} {\bibfnamefont {M.}~\bibnamefont
  {Rubinstein}}\ and\ \bibinfo {author} {\bibfnamefont {R.}~\bibnamefont
  {Colby}},\ }\href@noop {} {\emph {\bibinfo {title} {Polymer Physics}}}\
  (\bibinfo  {publisher} {Oxford University Press},\ \bibinfo {year}
  {2003})\BibitemShut {NoStop}%
\bibitem [{\citenamefont {Arora}\ \emph {et~al.}(2016)\citenamefont {Arora},
  \citenamefont {Qin}, \citenamefont {Morse}, \citenamefont {Delaney},
  \citenamefont {Fredrickson}, \citenamefont {Bates},\ and\ \citenamefont
  {Dorfman}}]{Arora2016}%
  \BibitemOpen
  \bibfield  {author} {\bibinfo {author} {\bibfnamefont {A.}~\bibnamefont
  {Arora}}, \bibinfo {author} {\bibfnamefont {J.}~\bibnamefont {Qin}}, \bibinfo
  {author} {\bibfnamefont {D.~C.}\ \bibnamefont {Morse}}, \bibinfo {author}
  {\bibfnamefont {K.~T.}\ \bibnamefont {Delaney}}, \bibinfo {author}
  {\bibfnamefont {G.~H.}\ \bibnamefont {Fredrickson}}, \bibinfo {author}
  {\bibfnamefont {F.~S.}\ \bibnamefont {Bates}},\ and\ \bibinfo {author}
  {\bibfnamefont {K.~D.}\ \bibnamefont {Dorfman}},\ }\bibfield  {title}
  {\bibinfo {title} {{Broadly Accessible Self-Consistent Field Theory for Block
  Polymer Materials Discovery}},\ }\href
  {https://doi.org/10.1021/acs.macromol.6b00107} {\bibfield  {journal}
  {\bibinfo  {journal} {Macromolecules}\ }\textbf {\bibinfo {volume} {49}},\
  \bibinfo {pages} {4675} (\bibinfo {year} {2016})}\BibitemShut {NoStop}%
\bibitem [{\citenamefont {Prasad}\ \emph {et~al.}(2018)\citenamefont {Prasad},
  \citenamefont {Jinnai}, \citenamefont {Ho}, \citenamefont {Thomas},\ and\
  \citenamefont {Grason}}]{Prasad2018}%
  \BibitemOpen
  \bibfield  {author} {\bibinfo {author} {\bibfnamefont {I.}~\bibnamefont
  {Prasad}}, \bibinfo {author} {\bibfnamefont {H.}~\bibnamefont {Jinnai}},
  \bibinfo {author} {\bibfnamefont {R.-M.}\ \bibnamefont {Ho}}, \bibinfo
  {author} {\bibfnamefont {E.~L.}\ \bibnamefont {Thomas}},\ and\ \bibinfo
  {author} {\bibfnamefont {G.~M.}\ \bibnamefont {Grason}},\ }\bibfield  {title}
  {\bibinfo {title} {Anatomy of triply-periodic network assemblies:
  characterizing skeletal and inter-domain surface geometry of block copolymer
  gyroids},\ }\href {https://doi.org/10.1039/C8SM00078F} {\bibfield  {journal}
  {\bibinfo  {journal} {Soft Matter}\ }\textbf {\bibinfo {volume} {14}},\
  \bibinfo {pages} {3612} (\bibinfo {year} {2018})}\BibitemShut {NoStop}%
\bibitem [{\citenamefont {Warner}\ and\ \citenamefont
  {Terentjev}(2007)}]{WarnerTerentjev2007}%
  \BibitemOpen
  \bibfield  {author} {\bibinfo {author} {\bibfnamefont {M.}~\bibnamefont
  {Warner}}\ and\ \bibinfo {author} {\bibfnamefont {E.}~\bibnamefont
  {Terentjev}},\ }\href@noop {} {\emph {\bibinfo {title} {Liquid Crystal
  Elastomers}}},\ International Series of Monographs on Physics\ (\bibinfo
  {publisher} {OUP Oxford},\ \bibinfo {year} {2007})\BibitemShut {NoStop}%
\bibitem [{\citenamefont {Conti}\ and\ \citenamefont
  {MacKintosh}(2009)}]{Conti2009}%
  \BibitemOpen
  \bibfield  {author} {\bibinfo {author} {\bibfnamefont {E.}~\bibnamefont
  {Conti}}\ and\ \bibinfo {author} {\bibfnamefont {F.~C.}\ \bibnamefont
  {MacKintosh}},\ }\bibfield  {title} {\bibinfo {title} {Cross-linked networks
  of stiff filaments exhibit negative normal stress},\ }\href
  {https://doi.org/10.1103/PhysRevLett.102.088102} {\bibfield  {journal}
  {\bibinfo  {journal} {Phys. Rev. Lett.}\ }\textbf {\bibinfo {volume} {102}},\
  \bibinfo {pages} {088102} (\bibinfo {year} {2009})}\BibitemShut {NoStop}%
\bibitem [{\citenamefont {Feng}\ \emph {et~al.}(2019)\citenamefont {Feng},
  \citenamefont {Burke}, \citenamefont {Zhou}, \citenamefont {Guo},
  \citenamefont {Yang}, \citenamefont {Reddy}, \citenamefont {Prasad},
  \citenamefont {Ho}, \citenamefont {Avgeropoulos}, \citenamefont {Grason},\
  and\ \citenamefont {Thomas}}]{Feng2019}%
  \BibitemOpen
  \bibfield  {author} {\bibinfo {author} {\bibfnamefont {X.}~\bibnamefont
  {Feng}}, \bibinfo {author} {\bibfnamefont {C.~J.}\ \bibnamefont {Burke}},
  \bibinfo {author} {\bibfnamefont {M.}~\bibnamefont {Zhou}}, \bibinfo {author}
  {\bibfnamefont {H.}~\bibnamefont {Guo}}, \bibinfo {author} {\bibfnamefont
  {K.}~\bibnamefont {Yang}}, \bibinfo {author} {\bibfnamefont {A.}~\bibnamefont
  {Reddy}}, \bibinfo {author} {\bibfnamefont {I.}~\bibnamefont {Prasad}},
  \bibinfo {author} {\bibfnamefont {R.-M.}\ \bibnamefont {Ho}}, \bibinfo
  {author} {\bibfnamefont {A.}~\bibnamefont {Avgeropoulos}}, \bibinfo {author}
  {\bibfnamefont {G.~M.}\ \bibnamefont {Grason}},\ and\ \bibinfo {author}
  {\bibfnamefont {E.~L.}\ \bibnamefont {Thomas}},\ }\bibfield  {title}
  {\bibinfo {title} {Seeing mesoatomic distortions in soft-matter crystals of a
  double-gyroid block copolymer},\ }\href
  {https://doi.org/10.1038/s41586-019-1706-1} {\bibfield  {journal} {\bibinfo
  {journal} {Nature}\ }\textbf {\bibinfo {volume} {575}},\ \bibinfo {pages}
  {175–179} (\bibinfo {year} {2019})}\BibitemShut {NoStop}%
\bibitem [{\citenamefont {Reddy}\ \emph {et~al.}(2022)\citenamefont {Reddy},
  \citenamefont {Dimitriyev},\ and\ \citenamefont {Grason}}]{Reddy2022}%
  \BibitemOpen
  \bibfield  {author} {\bibinfo {author} {\bibfnamefont {A.}~\bibnamefont
  {Reddy}}, \bibinfo {author} {\bibfnamefont {M.}~\bibnamefont {Dimitriyev}},\
  and\ \bibinfo {author} {\bibfnamefont {G.}~\bibnamefont {Grason}},\
  }\bibfield  {title} {\bibinfo {title} {Medial packing and elastic asymmetry
  stabilize the double-gyroid in block copolymers},\ }\bibfield  {journal}
  {\bibinfo  {journal} {Nature Communications}\ }\textbf {\bibinfo {volume}
  {13}},\ \href {https://doi.org/https://doi.org/10.1038/s41467-022-30343-2}
  {https://doi.org/10.1038/s41467-022-30343-2} (\bibinfo {year}
  {2022})\BibitemShut {NoStop}%
\bibitem [{\citenamefont {Dimitriyev}\ \emph {et~al.}(2023)\citenamefont
  {Dimitriyev}, \citenamefont {Reddy},\ and\ \citenamefont
  {Grason}}]{Dimitriyev2023}%
  \BibitemOpen
  \bibfield  {author} {\bibinfo {author} {\bibfnamefont {M.~S.}\ \bibnamefont
  {Dimitriyev}}, \bibinfo {author} {\bibfnamefont {A.}~\bibnamefont {Reddy}},\
  and\ \bibinfo {author} {\bibfnamefont {G.~M.}\ \bibnamefont {Grason}},\
  }\bibfield  {title} {\bibinfo {title} {Medial packing, frustration, and
  competing network phases in strongly segregated block copolymers},\ }\href
  {https://doi.org/10.1021/acs.macromol.3c01098} {\bibfield  {journal}
  {\bibinfo  {journal} {Macromolecules}\ }\textbf {\bibinfo {volume} {56}},\
  \bibinfo {pages} {7184} (\bibinfo {year} {2023})}\BibitemShut {NoStop}%
\bibitem [{\citenamefont {Wells}(1977)}]{Wells1977}%
  \BibitemOpen
  \bibfield  {author} {\bibinfo {author} {\bibfnamefont {A.~F.}\ \bibnamefont
  {Wells}},\ }\href@noop {} {\emph {\bibinfo {title} {Three Dimensional Nets
  and Polyhedra}}}\ (\bibinfo  {publisher} {Wiley},\ \bibinfo {address} {New
  York},\ \bibinfo {year} {1977})\BibitemShut {NoStop}%
\bibitem [{\citenamefont {Bailey}\ \emph {et~al.}(2002)\citenamefont {Bailey},
  \citenamefont {Hardy}, \citenamefont {Epps},\ and\ \citenamefont
  {Bates}}]{Bailey2002}%
  \BibitemOpen
  \bibfield  {author} {\bibinfo {author} {\bibfnamefont {T.~S.}\ \bibnamefont
  {Bailey}}, \bibinfo {author} {\bibfnamefont {C.~M.}\ \bibnamefont {Hardy}},
  \bibinfo {author} {\bibfnamefont {T.~H.}\ \bibnamefont {Epps}},\ and\
  \bibinfo {author} {\bibfnamefont {F.~S.}\ \bibnamefont {Bates}},\ }\bibfield
  {title} {\bibinfo {title} {A noncubic triply periodic network morphology in
  poly(isoprene-b-styrene-b-ethylene oxide) triblock copolymers},\ }\href
  {https://doi.org/10.1021/ma011716x} {\bibfield  {journal} {\bibinfo
  {journal} {Macromolecules}\ }\textbf {\bibinfo {volume} {35}},\ \bibinfo
  {pages} {7007} (\bibinfo {year} {2002})}\BibitemShut {NoStop}%
\bibitem [{\citenamefont {Tyler}\ and\ \citenamefont
  {Morse}(2005)}]{Tyler2005}%
  \BibitemOpen
  \bibfield  {author} {\bibinfo {author} {\bibfnamefont {C.~A.}\ \bibnamefont
  {Tyler}}\ and\ \bibinfo {author} {\bibfnamefont {D.~C.}\ \bibnamefont
  {Morse}},\ }\bibfield  {title} {\bibinfo {title} {Orthorhombic $fddd$ network
  in triblock and diblock copolymer melts},\ }\href
  {https://doi.org/10.1103/PhysRevLett.94.208302} {\bibfield  {journal}
  {\bibinfo  {journal} {Phys. Rev. Lett.}\ }\textbf {\bibinfo {volume} {94}},\
  \bibinfo {pages} {208302} (\bibinfo {year} {2005})}\BibitemShut {NoStop}%
\end{thebibliography}%

\end{document}



\title{Supplementary Information: \\Non-affinity of liquid networks and bicontinuous mesophases}

\author{Michael S.~Dimitriyev}
 \email{msdim@tamu.edu}
\affiliation{%
 Department of Polymer Science and Engineering, University of Massachusetts, Amherst, MA 01003, USA
}
\affiliation{%
 Department of Materials Science and Engineering, Texas A\&M University, College Station, TX 77843
}

\author{Xueyan Feng}
\affiliation{%
 Department of Macromolecular Science, State Key Laboratory of Molecular Engineering of Polymers, Fudan University, Shanghai, China 200437
}

\author{Edwin L.~Thomas}
\affiliation{%
 Department of Materials Science and Engineering, Texas A\&M University, College Station, TX 77843
}

\author{Gregory M.~Grason}%
 \email{grason@umass.edu}
\affiliation{%
 Department of Polymer Science and Engineering, University of Massachusetts, Amherst, MA 01003, USA
}

\renewcommand\thefigure{S\arabic{figure}}
\renewcommand\thesection{S\arabic{section}}
\renewcommand\theequation{S\arabic{equation}}

\maketitle

\section{Non-affinity of pre-stressed, equilibrium networks}

We start with nodes in a reference configuration specified by positions $\{\mathbf{r}^{(m)}\}$, assumed to be in an equilibrium configuration so that $\delta H[\{\mathbf{r}^{(m)}\}] = 0$.
Moreover, we assume that every strut in this configuration has length $\ell_0 = |\mathbf{r}^{(n)} - \mathbf{r}^{(m)}|$ for all node neighbors $m$ and $n$.
After an affine deformation, specified by deformation matrix $\Lambda$, the node positions are given by $\{\Lambda \mathbf{r}^{(m)}\}$.
The energy of this configuration is given by $H\left[\left\{\Lambda \mathbf{r}^{(m)}\right\}\right]$.
Next, the network will relax to a new equilibrium configuration.
Without some constraints on node positions, the network would simply relax to its non-deformed configuration.
These constraints come in two forms: (i) specification of ``boundary'' nodes that do not relax~\cite{Ivanov1994MinimalNT}; (ii) specification of fixed lattice vectors that constrain relate the positions of nodes in all copies of the unit cells~\cite{Alex2019,Alex2023}.
In the following we will only consider nodes that relax (i.e.~non-boundary nodes) with the idea that fixed nodes impose boundary conditions to the resulting equations.
To obtain the equilibrium configuration, consider the change in energy $\Delta H$ between an equilibrium configuration and the affine-deformed configuration, given by
\begin{equation}\begin{split}
\Delta H &\equiv H\left[\left\{\Lambda \mathbf{r}^{(m)} + \mathbf{u}^{(m)}\right\}\right] - H\left[\left\{\Lambda \mathbf{r}^{(m)}\right\}\right] \\
&= \sum_{mn}A_{mn}\bigg\{h\left(\left|\Lambda \mathbf{r}^{(n)} + \mathbf{u}^{(n)} - \Lambda \mathbf{r}^{(m)} - \mathbf{u}^{(m)}\right|\right) - h\left(\left|\Lambda \mathbf{r}^{(n)} - \Lambda \mathbf{r}^{(m)}\right|\right)\bigg\} \, ,
\end{split}\end{equation}
where relaxed node positions are displaced by vectors $\{\mathbf{u}^{(m)}\}$ from affine-deformed node positions.
Next, we expand $\Delta H$ to quadratic order in displacement:
\begin{equation}\begin{split}
\Delta H = \sum_{mn}A_{mn}\bigg\{& \frac{1}{2}(\mathbf{u}^{(m)} - \mathbf{u}^{(n)})^T\mathbf{D}^{(mn)}\,(\mathbf{u}^{(m)} - \mathbf{u}^{(n)})\\
&- \mathbf{f}^{(mn)}(\Lambda)\cdot(\mathbf{u}^{(m)} - \mathbf{u}^{(n)}) + \mathcal{O}(u^3) \bigg\}\, ,
\end{split}\end{equation}
where $\mathbf{D}^{(mn)}$ is a ``nodal'' dynamical matrix that weighs the cost of extending the strut joining node $m$ to $n$ in various directions and $\mathbf{f}^{(mn)}(\Lambda)$ is a force acting along that strut, which depends on the applied affine deformation $\Lambda$ of the unit cell.
Since each node of DP has pairs of anti-parallel struts, the net force on a given node is 0, so DP has a completely affine response.

In the linear-response limit, we expand $\Delta H$ to leading order in the symmetric strain
\begin{equation}
\bm{\varepsilon} \equiv \frac{1}{2}\left(\bm{\Lambda}^T\bm{\Lambda} - \mathbbm{1}\right)
\end{equation}
and find that the dynamical matrix is given by
\begin{equation}
\mathbf{D}^{(mn)} \approx \frac{\tau(\ell_0)}{\ell_0}\left(\mathbbm{1} - (1-\eta)\hat{\mathbf{R}}^{(mn)}\otimes\hat{\mathbf{R}}^{(mn)}\right)
\end{equation}
and the strut force is given by
\begin{equation}
\mathbf{f}^{(mn)} \approx \tau(\ell_0)\,(1-\eta)\, \varepsilon^{(mn)}\,\hat{\mathbf{R}}^{(mn)}
\end{equation}
where $\tau(\ell_0) = h'(\ell_0)$ is the residual tension of the struts in the undeformed configuration, $\varepsilon^{(mn)} \equiv \hat{R}^{(mn)}_i\, \varepsilon_{ij}\,\hat{R}^{(mn)}_j$ is a projection of the strain tensor onto the strut joining nodes $m$ and $n$, and $\eta \equiv \frac{\ell_0\,\tau'(\ell_0)}{\tau(\ell_0)}$ is a parameter that characterizes the tension-length constitutive relation.
Finally, the linear response $u \sim \varepsilon$ can be determined by minimizing $\Delta H$ with respect to the non-affine displacements of each node within the unit cell of a given network.
This yields, for each node $m$, an equilibrium condition,
\begin{equation}\label{eq:equilibrium_equations}
    \sum_{n \in \mathcal{N}_m}\left[ \mathbf{D}^{(mn)}(\mathbf{u}^{(m)} - \mathbf{u}^{(n)}) - \mathbf{f}^{(mn)} \right] = 0 \, ,
\end{equation}
where $\mathcal{N}_m$ is the set of neighbors of node $m$.
Given that the nodal dynamical matrix $\mathbf{D}$ and the affine force $\mathbf{f}$ have the residual stress $\tau(\ell_0)$ as an overall scale factor, we find that (i) the following results only hold for residually stressed networks (for which $\tau(\ell_0) \neq 0$) and (ii) the non-affine response is completely characterized by the parameter $\eta$.

\subsection{Trihedral and tetrahedral graphs}

First we solve Eq.~\ref{eq:equilibrium_equations} for the case of the trihedral and tetrahedral graphs shown in Fig.~2 of the main text.
In these cases, there is only one central node that relaxes to a new equilibrium position and the remaining 3 (trihedral) and 4 (tetrahedral) nodes are ``boundary nodes.''
It is useful to introduce coordinates to the nodes.
The central node $\mathbf{r}^{(0)}$ will be located at the origin for both graphs.
For the trihedral graph, the boundary nodes will be located at
\begin{equation}\begin{split}
    \mathbf{r}^{(1)} &= \frac{\ell_0}{\sqrt{2}} \left(1,\, -1,\, 0\right) \\
    \mathbf{r}^{(2)} &= \frac{\ell_0}{\sqrt{2}} \left(0,\, 1,\, -1\right) \\
    \mathbf{r}^{(3)} &= \frac{\ell_0}{\sqrt{2}} \left(-1,\, 0,\, 1\right) 
\end{split}\end{equation}
For the tetrahedral graph, the boundary nodes will be located at
\begin{equation}\begin{split}
    \mathbf{r}^{(1)} &= \frac{\ell_0}{\sqrt{3}} \left(1,\, 1,\, 1\right) \\
    \mathbf{r}^{(2)} &= \frac{\ell_0}{\sqrt{3}} \left(1,\, -1,\, -1\right) \\
    \mathbf{r}^{(3)} &= \frac{\ell_0}{\sqrt{3}} \left(-1,\, 1,\, -1\right) \\
    \mathbf{r}^{(4)} &= \frac{\ell_0}{\sqrt{3}} \left(-1,\, -1,\, 1\right)
\end{split}\end{equation}
For both graphs, we can rewrite Eq.~\ref{eq:equilibrium_equations} for the central node as
\begin{equation}
    \mathbf{D}\,\mathbf{u} = \mathbf{f} \, ,
\end{equation}
where we suppress the superscript on the displacement vector $\mathbf{u} \equiv \mathbf{u}^{(0)}$.
Then the scale of the non-affine displacement is given by
\begin{equation}
\frac{|\mathbf{u}|^2}{\ell_0^2} = \frac{1}{\ell_0^2}\mathbf{f}^T(\mathbf{D}^{-1})^T\mathbf{D}^{-1}\mathbf{f} \, .
\end{equation}
For the trihedral network, the dynamical matrix $\mathbf{D}$ is given by
\begin{equation}\begin{split}
\mathbf{D} &= \frac{3\tau(\ell_0)(1 + \eta)}{2 \ell_0}\left\{\mathbbm{1} + \frac{1- \eta}{1 + \eta}\hat{\mathbf{N}}\otimes\hat{\mathbf{N}}\right\}
\end{split}\end{equation}
where $\hat{\mathbf{N}}$ is the direction normal to the plane of the trihedral graph (here, $\hat{\mathbf{N}} = (1,1,1)/\sqrt{3}$).
For the tetrahedral network, the dynamical matrix is given by
\begin{equation}
\mathbf{D} = \frac{8\tau(\ell_0)(1+\frac{\eta}{2})}{3\ell_0}\mathbbm{1} \, .
\end{equation}
We will focus on the case of uniaxial deformation, where
\begin{equation}
\bm{\varepsilon} = \varepsilon\, \hat{\mathbf{d}}\otimes \hat{\mathbf{d}} \, .
\end{equation}
Consequently, for the trihedral graph,
\begin{equation}
    \frac{|\mathbf{u}|^2}{\ell_0^2} = \frac{1}{4}\left(\frac{1-\eta}{1+\eta}\right)^2\,|\hat{\mathbf{d}}\times\hat{\mathbf{N}}|^4\,\varepsilon^2 = \Gamma_{\rm tri.}(\hat{\mathbf{d}})\,\varepsilon^2
\end{equation}
and for the tetrahedral graph,
\begin{equation}
    \frac{|\mathbf{u}|^2}{\ell_0^2} = \frac{1}{6}\left(\frac{1-\eta}{1+\frac{\eta}{2}}\right)^2\,\left[1 - \hat{d}_x^4 -\hat{d}_y^4 -\hat{d}_z^4\right]\,\varepsilon^2 = \Gamma_{\rm tet.}(\hat{\mathbf{d}})\,\varepsilon^2 \, .
\end{equation}
Writing $\Gamma(\hat{\mathbf{d}}) = \Gamma^{\rm max}\psi(\hat{\mathbf{d}})$, we have
\begin{equation}\begin{split}\label{eq:tri_result}
    \Gamma^{\rm tri.}_{\rm max} &= \frac{1}{4}\left(\frac{1-\eta}{1+\eta}\right)^2 \\
    \psi^{\rm tri.}(\hat{\mathbf{d}}) &= |\hat{\mathbf{d}}\times\hat{\mathbf{N}}|^4
\end{split}\end{equation}
and
\begin{equation}\begin{split}\label{eq:tet_result}
    \Gamma^{\rm tet.}_{\rm max} &= \frac{1}{9}\left(\frac{1-\eta}{1+\frac{\eta}{2}}\right)^2 \\
    \psi^{\rm tet.}(\hat{\mathbf{d}}) &= \frac{3}{2}\left[1 - \hat{d}_x^4 -\hat{d}_y^4 -\hat{d}_z^4\right]
\end{split}\end{equation}
Finally, we determine the average $\overline{\Gamma}$ by integrating $\Gamma(\hat{\mathbf{d}})$ over the 2-sphere,
\begin{equation}
    \overline{\Gamma} =  \Gamma_{\rm max}\oint_{\mathbbm{S}^2} \frac{{\rm d}\Omega}{4\pi}\, \psi(\hat{\mathbf{d}})\, ,
\end{equation}
yielding
\begin{equation}\begin{split}
\overline{\Gamma}^{\rm tri.} &= \frac{8}{15}\Gamma^{\rm tri.}_{\rm max} \\
\overline{\Gamma}^{\rm tet.} &= \frac{2}{5}\Gamma^{\rm tet.}_{\rm max}
\end{split}\end{equation}

\subsection{DG and DD networks}

Triply-periodic networks do not possess boundary nodes; instead, network equilibrium is determined by the position of nodes within a given unit cell.
In this case, Eq.~\ref{eq:equilibrium_equations} must be solved with periodic boundary conditions.
A single DG network has initial node positions given by
\begin{equation}\begin{split}
\mathbf{r}^{(1)} &= \frac{1}{8}\left(1,1,1\right) \\
\mathbf{r}^{(2)} &= \frac{1}{8}\left(3,7,5\right) \\
\mathbf{r}^{(3)} &= \frac{1}{8}\left(7,5,3\right) \\
\mathbf{r}^{(4)} &= \frac{1}{8}\left(5,3,7\right) \\
\mathbf{r}^{(5)} &= \frac{1}{8}\left(5,5,5\right) \\
\mathbf{r}^{(6)} &= \frac{1}{8}\left(1,7,3\right) \\
\mathbf{r}^{(7)} &= \frac{1}{8}\left(7,3,1\right) \\
\mathbf{r}^{(8)} &= \frac{1}{8}\left(3,1,7\right) 
\end{split}\end{equation}
and edge relations given by
\begin{equation}\begin{split}
\langle 1,6 \rangle &: \mathbf{r}^{(1)} \to \mathbf{r}^{(6)} + (0, -1, 0) \\
\langle 1,7 \rangle &: \mathbf{r}^{(1)} \to \mathbf{r}^{(7)} + (-1, 0, 0) \\
\langle 1,8 \rangle &: \mathbf{r}^{(1)} \to \mathbf{r}^{(6)} + (0, 0, -1) \\
\langle 2,5 \rangle &: \mathbf{r}^{(2)} \to \mathbf{r}^{(5)} + (0, 0, 0) \\
\langle 2,6 \rangle &: \mathbf{r}^{(2)} \to \mathbf{r}^{(6)} + (0, 0, 0) \\
\langle 2,8 \rangle &: \mathbf{r}^{(2)} \to \mathbf{r}^{(8)} + (0, 1, 0) \\
\langle 3,5 \rangle &: \mathbf{r}^{(3)} \to \mathbf{r}^{(5)} + (0, 0, 0) \\
\langle 3,6 \rangle &: \mathbf{r}^{(3)} \to \mathbf{r}^{(6)} + (1, 0, 0) \\
\langle 3,7 \rangle &: \mathbf{r}^{(3)} \to \mathbf{r}^{(7)} + (0, 0, 0) \\
\langle 4,5 \rangle &: \mathbf{r}^{(4)} \to \mathbf{r}^{(5)} + (0, 0, 0) \\
\langle 4,7 \rangle &: \mathbf{r}^{(4)} \to \mathbf{r}^{(7)} + (0, 0, 1) \\
\langle 4,8 \rangle &: \mathbf{r}^{(4)} \to \mathbf{r}^{(8)} + (0, 0, 0) 
\end{split}\end{equation}
where $\langle mn\rangle$ denotes a pair of joined nodes, with strut joining node $\mathbf{r}^{(m)}$ to $\mathbf{r}^{(n)} + \Delta\mathbf{e}^{(mn)}$, where $\Delta\mathbf{e}^{(mn)}$ is an additional translation vector that accounts for struts that pass through a unit cell boundary.
Similarly, a single DD network has initial node positions given by
\begin{equation}\begin{split}
\mathbf{r}^{(1)} &= \frac{1}{8}\left(1,1,1\right) \\
\mathbf{r}^{(2)} &= \frac{1}{8}\left(3,3,3\right) \\
\mathbf{r}^{(3)} &= \frac{1}{8}\left(1,5,5\right) \\
\mathbf{r}^{(4)} &= \frac{1}{8}\left(5,1,5\right) \\
\mathbf{r}^{(5)} &= \frac{1}{8}\left(5,5,1\right) \\
\mathbf{r}^{(6)} &= \frac{1}{8}\left(3,7,7\right) \\
\mathbf{r}^{(7)} &= \frac{1}{8}\left(7,3,7\right) \\
\mathbf{r}^{(8)} &= \frac{1}{8}\left(7,7,3\right) 
\end{split}\end{equation}
and edge relations given by
\begin{equation}\begin{split}
\langle 1,2 \rangle &: \mathbf{r}^{(1)} \to \mathbf{r}^{(2)} + (0, 0, 0) \\
\langle 1,6 \rangle &: \mathbf{r}^{(1)} \to \mathbf{r}^{(6)} + (0, -1, -1) \\
\langle 1,7 \rangle &: \mathbf{r}^{(1)} \to \mathbf{r}^{(7)} + (-1, 0, -1) \\
\langle 1,8 \rangle &: \mathbf{r}^{(1)} \to \mathbf{r}^{(8)} + (-1, -1, 0) \\
\langle 2,3 \rangle &: \mathbf{r}^{(2)} \to \mathbf{r}^{(3)} + (0, 0, 0) \\
\langle 2,4 \rangle &: \mathbf{r}^{(2)} \to \mathbf{r}^{(4)} + (0, 0, 0) \\
\langle 2,5 \rangle &: \mathbf{r}^{(2)} \to \mathbf{r}^{(5)} + (0, 0, 0) \\
\langle 3,6 \rangle &: \mathbf{r}^{(3)} \to \mathbf{r}^{(6)} + (0, 0, 0) \\
\langle 3,7 \rangle &: \mathbf{r}^{(3)} \to \mathbf{r}^{(7)} + (-1, 0, 0) \\
\langle 3,8 \rangle &: \mathbf{r}^{(3)} \to \mathbf{r}^{(8)} + (-1, 0, 0) \\
\langle 4,6 \rangle &: \mathbf{r}^{(4)} \to \mathbf{r}^{(6)} + (0, -1, 0) \\
\langle 4,7 \rangle &: \mathbf{r}^{(4)} \to \mathbf{r}^{(7)} + (0, 0, 0) \\
\langle 4,8 \rangle &: \mathbf{r}^{(4)} \to \mathbf{r}^{(8)} + (0, -1, 0) \\
\langle 5,6 \rangle &: \mathbf{r}^{(5)} \to \mathbf{r}^{(6)} + (0, 0, -1) \\
\langle 5,7 \rangle &: \mathbf{r}^{(5)} \to \mathbf{r}^{(7)} + (0, 0, -1) \\
\langle 5,8 \rangle &: \mathbf{r}^{(5)} \to \mathbf{r}^{(8)} + (0, 0, 0) \\
\end{split}\end{equation}
Note that the DD skeletal graph detailed above is generated from the symmetry group $Fd\overline{3}m$ and contains 8 copies of the standard DD unit cell generated from the $Pn\overline{3}m$ space group.
The above form allows for a straightforward representation of single networks, similar to the representation used for a the DG skeletal graph.

Next, it is convenient to represent $\{\mathbf{u}^{(m)}\}$ as a $3 N_{\rm nodes}$-dimensional vector
\begin{equation}
    \mathbf{U} = \left(\begin{array}{c}
    u^{(1)}_x \\
    u^{(1)}_y \\
    u^{(1)}_z \\
    u^{(2)}_x \\
    \vdots \\
    u^{(N_{\rm nodes})}_z
    \end{array}\right),
\end{equation}
where $N_{\rm nodes}$ is the number of nodes per unit cell (8 for each of the single networks above).
Each $3\times 3$ nodal dynamical matrix $\mathbf{D}^{(mn)}$ can then be folded into a $3 N_{\rm nodes} \times 3 N_{\rm nodes}$ dynamical matrix $\mathbbm{D}$ and the sum over forces at each node, $\sum_{n\in \langle mn \rangle}\mathbf{f}^{(mn)}$, can be written as a $3 N_{\rm nodes}$-dimensional vector $\mathbf{F}$.
As a result, the equilibrium node displacement vector $\mathbf{U}$ satisfies a modified form of Eq.~\ref{eq:equilibrium_equations}, namely
\begin{equation}\label{eq:modified_equilibrium_equations}
    \mathbbm{D}\mathbf{U} = \mathbf{F} \, .
\end{equation}
Since the energy of the network is invariant under continuous, rigid translations (i.e., $\mathbf{u}^{(m)} = {\rm const.}$ for all $m$), the dynamical matrix $\mathbbm{D}$ possesses three zero-eigenvalues and is therefore not invertible.
To solve Eq.~\ref{eq:modified_equilibrium_equations}, we restrict the possible solution set of $\mathbf{U}$ to exclude rigid translations.
We do this via the \emph{Moore-Penrose pseudoinverse} $\mathbbm{D}^+$, which acts as the inverse of $\mathbbm{D}$ in the complement to its nullspace.
Thus, solutions are given by
\begin{equation}
    \mathbf{U} = \mathbbm{D}^+ \mathbf{F}
\end{equation}
and the non-affine response is given by
\begin{equation}\label{eq:na_response_general}
    \frac{\langle |\mathbf{u}|^2\rangle}{\ell_0^2} = \frac{\left|\mathbf{U}\right|^2}{N_{\rm nodes}\ell_0^2} = \frac{1}{N_{\rm nodes}{\ell_0^2}}\mathbf{U}^{T}\left(\mathbbm{D}^+\right)^{T}\mathbbm{D}^+\mathbf{U} \, .
\end{equation}
We use Mathematica to compute the pseudoinverse $\mathbbm{D}^+$ and the non-affine response to strain components $\varepsilon_{ij}$ via Eq.~\ref{eq:na_response_general}, yielding
\begin{equation}\begin{split}
\frac{\langle |\mathbf{u}|^2\rangle_{\rm DG}}{\ell_0^2} &= \frac{3}{12}\left(\frac{1-\eta}{1 + 3\eta}\right)^2\left[\varepsilon_{xx}^2 + \varepsilon_{yy}^2 + \varepsilon_{zz}^2 - \varepsilon_{xx}\varepsilon_{yy} - \varepsilon_{yy}\varepsilon_{zz} - \varepsilon_{zz}\varepsilon_{xx}\right] + \frac{1}{4}\left(\frac{1-\eta}{1 + \eta}\right)^2\left[\varepsilon_{xy}^2 + \varepsilon_{yz}^2 + \varepsilon_{zx}^2\right]\\
\frac{\langle |\mathbf{u}|^2\rangle_{\rm DD}}{\ell_0^2} &= \frac{1}{3}\left(\frac{1 - \eta}{2 + \eta}\right)^2\left[\varepsilon_{xy}^2 + \varepsilon_{yz}^2 + \varepsilon_{zx}^2\right]
\end{split}\end{equation}
from which we determine the non-affinity components.

\section{Comparisons with strain-stiffening polymers}

\begin{figure}[h!]
    \centering
    \includegraphics[width=.6\columnwidth]{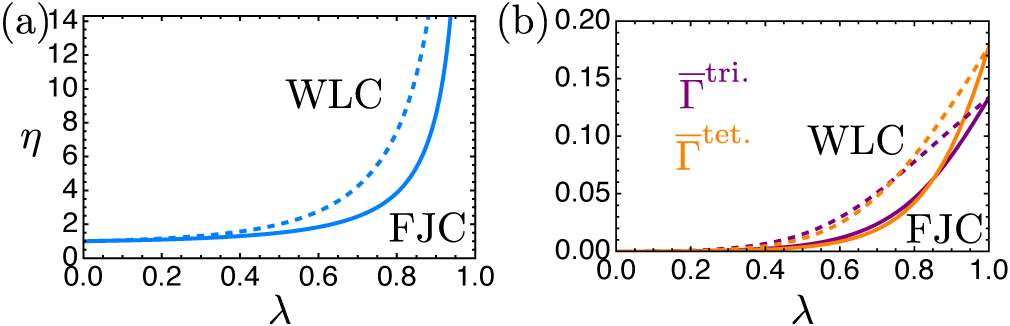}
    \caption{\label{fig:FENE} (a) Effective force law parameter $\eta$ for the Freely-Jointed Chain (FJC) model (solid) and the Worm-Like Chain (WLC) model (dashed) as a function of chain stretch relative to chain length $\lambda$. (b) Non-affinity estimates from linear response theory for the trihedral graph (purple) and the tetrahedral graph (orange) for both chain models.}
\end{figure}

Polymer networks strain-stiffen due to the finite extensibility of polymer chains when strongly stretched.
Two models that describe polymers of finite extent are the Freely-Jointed Chain (FJC) model of flexible polymers and the Worm-Like Chain (WLC) model~\cite{Marko1995,Rubinstein2003}.
The force $f$ needed to stretch a chain to end-separation $r$ is well-approximated within both models as
\begin{equation}
    \frac{f b}{k_{\rm B} T} = \frac{2 r}{N b} + \frac{1}{p}\left(1 - \frac{r}{Nb}\right)^{-p} - \frac{1}{p} \, ,
\end{equation}
where $b$ is the statistical segment length, $N$ is the polymerization index, $k_{\rm B} T$ is the thermal energy scale, and $p = 1$ (FJM) or 2 (WLC).
Using these force laws as constitutive models for strut tension, we can estimate the non-affine response of finitely-extensible polymer networks from the characteristic parameter $\eta(r) = r f'/f$.
Defining $\lambda \equiv r/(N b)$ as the stretch fraction relative to total chain length, the characteristic parameter is given by
\begin{equation}
    \eta(\lambda) = \frac{1 + \frac{1}{2}\left(1 - \lambda\right)^{-p-1}}{1 + \frac{1}{2 p \lambda}\left[\left(1 - \lambda\right)^{-p} - 1\right]} \, ,
\end{equation}
which only depends on $\lambda$ and $p$.
As shown in Fig.~\ref{fig:FENE}(a), the parameter $\eta \geq 1$, correctly categorizing these models as strain-stiffening; as $\lambda \to 1$, $\eta \to \infty$.
Consequently, as $\lambda \to 1$, the expected non-affine response approaches values similar to those expected by LNT, as shown in Fig.~\ref{fig:FENE}(b).

\section{Symmetries of linear response}

The non-affinity tensor $\bm{\Gamma}$ provides the quadratic-order strain coupling in the linear-response regime of deformation.
Much like the elastic modulus tensor of linear elasticity, $\bm{\Gamma}$ possesses a variety of symmetries, including
\begin{itemize}
\item Minor symmetries: $\Gamma^{ijkl} = \Gamma^{jikl} = \Gamma^{ijlk}$
\item Major symmetry: $\Gamma^{ijkl} = \Gamma^{klij}$
\item Symmetries related to the crystal system
\end{itemize}
Due to the symmetry of the strain tensor $\varepsilon_{ij} = \varepsilon_{ji}$ and the minor symmetries of $\bm{\Gamma}$, it is useful to represent both using Voigt notation $\varepsilon_{ij} \mapsto \varepsilon_{I}$ and $\Gamma^{ijkl} \mapsto \Gamma^{IJ}$, where $I,J,\dots \in \{1,2,\dots,6\}$ represents, in order, the set of indices $\{xx, yy, zz, yz, xz, xy\}$.
Setting $\{\varepsilon_1, \varepsilon_2, \varepsilon_3\} = \{\varepsilon_{xx}, \varepsilon_{yy}, \varepsilon_{zz}\}$ and $\{\varepsilon_4, \varepsilon_5, \varepsilon_6\} = \{2\varepsilon_{yz}, 2\varepsilon_{xz}, 2\varepsilon_{xy}\}$ allows for the identity $\langle |\mathbf{u}|^2 \rangle/\ell_0^2 = \Gamma^{ijkl}\varepsilon_{ij}\varepsilon_{kl} = \Gamma^{IJ}\varepsilon_I \varepsilon_J$.
The cubic crystal system is invariant under $\pi/2$ rotations of each axis; much like the elastic modulus tensor, $\Gamma$ thus consists of three non-zero, independent constants (reduced from 21 for a full triclinic system),
\begin{equation}
    \Gamma^{IJ} = \left\{\begin{array}{ccc|ccc}
    A & B & B & 0 & 0 & 0 \\
    B & A & B & 0 & 0 & 0 \\
    B & B & A & 0 & 0 & 0 \\
    \hline
    0 & 0 & 0 & C & 0 & 0 \\
    0 & 0 & 0 & 0 & C & 0 \\
    0 & 0 & 0 & 0 & 0 & C
    \end{array}\right\}
\end{equation}
where, in order for $\langle |\mathbf{u}|^2 \rangle$ to be positive semidefinite ($\langle |\mathbf{u}|^2 \rangle \geq 0$), the constants $A$, $B$, and $C$ must be such that ${\rm det}\,\bm{\Gamma} = (A - B)^2(A + 2B)C^3 \geq 0$.
Thus, $\langle |\mathbf{u}|^2 \rangle$ is given by
\begin{equation}
    \frac{\langle |\mathbf{u}|^2\rangle}{\ell_0^2} = A\left(\varepsilon^2_{xx} + \varepsilon^2_{yy} + \varepsilon^2_{zz}\right) + 2B\left(\varepsilon_{xx}\varepsilon_{yy} + \varepsilon_{yy}\varepsilon_{zz} + \varepsilon_{zz}\varepsilon_{xx}\right) + 4C\left(\varepsilon_{xy}^2 + \varepsilon_{yz}^2 + \varepsilon_{zx}^2\right) \, .
\end{equation}
Defining the deviatoric strain tensor (the traceless part of the strain tensor) as
\begin{equation}
\hat{\varepsilon}_{ij} \equiv \varepsilon_{ij} - \frac{1}{3}\delta_{ij}\varepsilon_{kk} \, ,
\end{equation}
we have
\begin{equation}
    \frac{\langle |\mathbf{u}|^2\rangle}{\ell_0^2} = A\left(\hat{\varepsilon}^2_{xx} + \hat{\varepsilon}^2_{yy} + \hat{\varepsilon}^2_{zz}\right) + 2B\left(\hat{\varepsilon}_{xx}\hat{\varepsilon}_{yy} + \hat{\varepsilon}_{yy}\hat{\varepsilon}_{zz} + \hat{\varepsilon}_{zz}\hat{\varepsilon}_{xx}\right) + \frac{A + 2B}{3}\left({\rm tr}\,\bm{\varepsilon}\right)^2 + 4C\left(\varepsilon_{xy}^2 + \varepsilon_{yz}^2 + \varepsilon_{zx}^2\right) \, .
\end{equation}
For the class of mechanical networks considered here, a force-balance is unchanged under a uniform re-scaling of space; consequently, $A + 2B = 0$.
The fact that this relationship implies that ${\rm det}\,\bm{\Gamma} = 0$ merely reflects the dilational symmetry of such mechanical networks.
Thus, we can write
\begin{equation}\begin{split}
    \frac{\langle |\mathbf{u}|^2\rangle}{\ell_0^2} &= \frac{A}{2}\left((\hat{\varepsilon}_{xx} - \hat{\varepsilon}_{yy})^2 + (\hat{\varepsilon}_{yy} - \hat{\varepsilon}_{zz})^2 + (\hat{\varepsilon}_{zz} - \hat{\varepsilon}_{xx})^2\right) + 4C\left(\varepsilon_{xy}^2 + \varepsilon_{yz}^2 + \varepsilon_{zx}^2\right) \\
    &= 3A\sum_i \frac{\hat{\varepsilon}_{ii}^2}{2} + 2C\sum_{i\neq j}\varepsilon_{ij}^2
\end{split}\end{equation}
and identify the first term as relating to volume-preserving extensional strain $\varepsilon_{\rm ext}$, where
\begin{equation}
\varepsilon_{\rm ext} \equiv \sqrt{\sum_i \frac{\hat{\varepsilon}_{ii}^2}{2}} = \sqrt{\frac{\hat{\varepsilon}_{xx}^2 + \hat{\varepsilon}_{yy}^2 + \hat{\varepsilon}_{zz}^2}{2}} \Bigg(= \sqrt{\frac{(\hat{\varepsilon}_{xx} - \hat{\varepsilon}_{yy})^2 + (\hat{\varepsilon}_{yy} - \hat{\varepsilon}_{zz})^2 + (\hat{\varepsilon}_{zz} - \hat{\varepsilon}_{xx})^2}{6}}\Bigg)
\end{equation}
(here, equality with the term in the parenthesis is ensured since ${\rm tr}\,\hat{\bm{\varepsilon}} = 0$) and the second term as relating to the shear strain $\varepsilon_{\rm shear}$, where
\begin{equation}
\varepsilon_{\rm shear} \equiv \sqrt{\sum_{i\neq j}\varepsilon_{ij}^2} = \sqrt{2\left(\varepsilon_{xy}^2 + \varepsilon_{yz}^2 + \varepsilon_{zx}^2\right)} \, .
\end{equation}
Thus, we have $3A = \Gamma_{\rm ext}$ and $2C = \Gamma_{\rm shear}$.

Finally, to plot multiple strain paths on the same axis, we use the above result that to linear order, the response can be represented as a function of $\varepsilon_{\rm ext}$ and $\varepsilon_{\rm shear}$, to project the full deformation path on a linearized subspace spanned by $\{\varepsilon_{\rm ext}, \varepsilon_{\rm shear}\}$.
We can then introduce polar angle $\alpha$ via
\begin{equation}
    \tan \alpha \equiv \frac{\varepsilon_{\rm shear}}{\varepsilon_{\rm ext}}
\end{equation}
and radial measure $\varepsilon_r$ via
\begin{equation}
    \varepsilon_r^2 \equiv \varepsilon_{\rm ext}^2 + \varepsilon_{\rm shear}^2 = \sum_i \frac{\hat{\varepsilon}_{ii}^2}{2} + \sum_{i\neq j}\varepsilon_{ij}^2 = \frac{1}{2}{\rm tr}\left(\hat{\bm{\varepsilon}}^2\right) \, ,
\end{equation}
where the final term in the above equality is used to plot non-affinity, length variation, and volume variation data on a common axis.
Note that the above arguments showing that the linear non-affine response $\langle |\mathbf{u}|^2 \rangle$ depends only on volume-preserving extensional and shear strain measurements holds for the linear response of strut length variations $\langle \Delta \ell^2 \rangle$ and nodal volume variations $\langle \Delta V^2 \rangle$ as both measures are similarly expected to be invariant under uniform dilations.

\section{SCF procedure}

Self-consistent field (SCF) theory calculations were performed using the open-source PSCF software~\cite{Arora2016}.
Using parameters relevant to the experimental data, we set A-block fractions as $f_{\rm A} = 0.4$ for DG and $f_{\rm B} = 0.42$ for DD and assumed equal statistical segment lengths for both blocks.
Finally, we chose $\chi N = 30$ as representative of the intermediate-segregation regime.

In order to calculate equilibrium density fields along simulated deformation paths, we first calculate the equilibrium fields assuming a cubic unit cell and constraining Fourier modes to obey the relevant space group; unit cell volume was allowed to fluctuate.
Using the equilibrium unit cell dimensions and fields obtained from this method as input, we then relax the space group constraint, impose a fixed volume (determined from the result of the first calculation), and calculate the resulting equilibrium structure.
We then iteratively adjust the unit cell dimension (at fixed volume), and use the equilibrium fields produced by the prior calculation as a starting point for each new calculation.
In this way, we are able to sweep out a deformation path at fixed volume.

In general, the deformed, triclinic unit cell can be parameterized by 3 lattice constants $a$, $b$, and $c$, as well as 3 angles $\alpha$, $\beta$, and $\gamma$.
Adopting the parameterization used in PSCF, the each triclinic unit cell can be transformed from a cubic unit cell by the deformation matrix
\begin{equation}\label{eq:def_mat}
\bm{\Lambda} = \frac{1}{a_0}\left(\begin{array}{ccc}
a & 0 & 0 \\
b \cos \gamma & b \sin \gamma & 0 \\
c \cos \alpha \sin \beta & c \sin \alpha \sin \beta & c \cos \beta
\end{array}\right) \, ,
\end{equation}
where $a_0$ is the lattice constant of the (undeformed) cubic unit cell.
We parameterize extensional deformations are by the 1D family of deformation matrices $\bm{\Lambda}_{\rm ext}(s)$, where
\begin{equation}
\bm{\Lambda}_{\rm ext}(s) = \left(\begin{array}{ccc}
1 + s & 0 & 0 \\
0 & \frac{1}{\sqrt{1+s}} & 0 \\
0 & 0 & \frac{1}{\sqrt{1+s}}
\end{array}\right)
\end{equation}
and shear deformations by the 1D family $\bm{\Lambda}_{\rm shear}(s)$, where
\begin{equation}
\bm{\Lambda}_{\rm shear}(s) = \left(\begin{array}{ccc}
1 & 0 & 0 \\
s & 1 & 0 \\
0 & 0 & 1
\end{array}\right) \, .
\end{equation}
To linear order, $\varepsilon_{\rm ext} = \frac{\sqrt{3}}{2}s$ and $\varepsilon_{\rm shear} = 0$ for the extensional deformation and $\varepsilon_{\rm shear} = \frac{s}{\sqrt{2}}$ and $\varepsilon_{\rm ext} = \mathcal{O}(s^2)$ for the shear deformation.

\begin{figure}[h!]
    \centering
    \includegraphics[width=.5\textwidth]{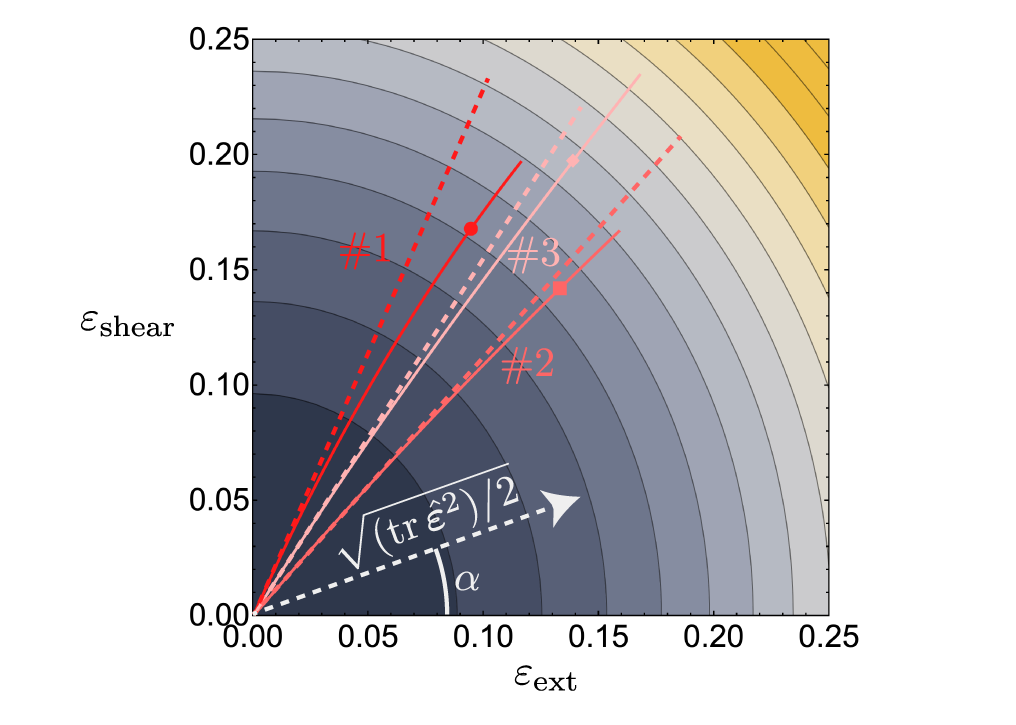}
    \caption{\label{fig:def_paths} Deformation paths \#1, \#2, and \#3 for DG (solid curves), projected onto the space spanned by the linear combination of extensional strain $\varepsilon_{\rm ext}$ and shear strain $\varepsilon_{\rm shear}$. These curves pass through points representing the three triclinic DG grains from experimental data. Dashed lines represent the linearization of these paths in the small-strain regime. Also shown is the polar representation of this space, where $\alpha$ is the polar angle and $\sqrt{({\rm tr}\,\hat{\bm{\varepsilon}}^2)/2} = \sqrt{\varepsilon_{\rm ext}^2 + \varepsilon_{\rm shear}^2}$ is the radial coordinate. The three linearized paths have polar angles $\alpha_1 \approx 66^\circ$, $\alpha_2 \approx 49^\circ$, and $\alpha_3 \approx 58^\circ$. Contours represent level sets of constant non-affine response $\langle |\mathbf{u}|^2 \rangle/\ell_0^2$, in the linear response approximation, as obtained from SCF measurements.}
\end{figure}

In order to simulate a more general, volume-preserving, deformation path, we take advantage of the exponential map
\begin{equation}
    \bm{\Lambda}(t) = e^{t\, \bm{\lambda}}
\end{equation}
for some generating matrix $\lambda$.
Due to the identity
\begin{equation}
    {\rm det}\,\bm{\Lambda}(t) = {\rm det}\,e^{t\, \bm{\lambda}} = e^{ t\, {\rm tr}\, \bm{\lambda}}\, ,
\end{equation}
it is guaranteed that if the generating matrix is traceless (i.e.~${\rm tr}\,\bm{\lambda} = 0$) then volume preserved at each point along the deformation (i.e.~${\rm det}\,\bm{\Lambda}(t) = 1$).
In order to determine an appropriate generating matrix $\bm{\lambda}$, we select a ``target'' deformation matrix $\bm{\Lambda}_1 = \bm{\Lambda}(t = 1)$.
Then the generating matrix is found by the logarithm map, $\bm{\lambda} = {\rm ln}\,\bm{\Lambda}_1$.
We apply this method to simulate deformation ``trajectories'' that originate at the cubic configuration (where $\bm{\Lambda}(0) = \mathbbm{1}$) and pass through the experimental point at $t = 1$ (where $\bm{\Lambda}(1) = \bm{\Lambda}_1$).
As shown in Fig.~\ref{fig:def_paths}, the resulting deformation paths are slightly curved, when projected onto the space spanned by the linear combination of extensional strain $\varepsilon_{\rm ext}$ and shear strain $\varepsilon_{\rm shear}$.
This curvature arises from non-linear strain coupling that occurs for large deformations.

\section{Measuring non-affine response from SCF calculations}

SCF calculations output density fields for the A- and B-block monomers.
Using these density fields and the skeletonization process outlined in ref.~\cite{Prasad2018}, we extract skeletal graphs for each relaxed network structure.
We then compare these skeletal graphs to affine deformations of the skeletal graphs of the cubic structures.
Each node $m$ in the skeletal relaxed graph can be obtained from a corresponding node in affinely-deformed skeletal graph by a displacement vector $\mathbf{u}^{(m)} = \mathbf{r}^{(m)} - \mathbf{r}_{\rm aff}^{(m)}$.
Given that the SCF calculations of the deformed networks involve the full set of Fourier modes within a single triclinic unit cell, periodic boundary conditions allow for free translation of the density field within the unit cell; we find that indeed there is a slight translation of the density field with respect to the usual choice of origin.
We therefore add a translation vector $\mathbf{t}$ so that the non-affine displacement vectors are given by
\begin{equation}
    \mathbf{u}^{(m)} = \mathbf{r}^{(m)} - (\mathbf{r}_{\rm aff}^{(m)} + \mathbf{t}) \, .
\end{equation}
Then we determine a measure of the total displacement by the average displaced distance,
\begin{equation}
    \langle |\mathbf{u}|^2 \rangle = \sum_{m = 1}^{N_{\rm nodes}} \frac{\left\vert \mathbf{u}^{(m)}\right\vert^2}{N_{\rm nodes}} \, ,
\end{equation}
where $N_{\rm nodes}$ is the number of nodes within the unit cell, corresponding to the multiplicity of the Wyckoff position: $N_{\rm nodes} = 16$ for DG, where nodes sit at the 16b Wyckoff positions; $N_{\rm nodes} = 2$ for DD, where nodes sit at the 2a Wyckoff positions.
Finally, we minimize $\langle |\mathbf{u}|^2 \rangle$ with respect to the translation vector $\mathbf{t}$ to ``subtract out'' the residual affine deformation that comes from translation, leaving a measure of non-affine response.

\subsection{Determination of non-affinity components}

In order to measure $\Gamma_{\rm ext}$ and $\Gamma_{\rm shear}$ from this process, we perform a pair of sweeps along deformation paths $\bm{\Lambda}_{\rm ext}(s)$ and $\bm{\Lambda}_{\rm shear}(s)$ involving (to linear order) only extensional strain and shear strain components, respectively.
Using the method outlined above, we extract a pair of non-affine responses, $\langle |\mathbf{u}|^2 \rangle_{\rm ext}/\ell_0^2$ and $\langle |\mathbf{u}|^2 \rangle_{\rm shear}/\ell_0^2$ as functions of individual strain components $\varepsilon_{\rm ext}$ and $\varepsilon_{\rm shear}$, respectively.
Next, we fit these data to functions of the form $\langle |\mathbf{u}|^2 \rangle/\ell_0^2 = \Gamma\,\varepsilon^2 + C$, where $C$ is an offset that represents the additional non-affine response originating from numerical error at $\varepsilon = 0$.
However, given that we only seek to quantify the \emph{linear response} (i.e.~the lowest-order, quadratic contribution to $\langle |\mathbf{u}|^2 \rangle$), the fit must be taken over a suitably chosen region of data.
This is achieved by choice of a cut-off strain $\varepsilon_{\rm cut}$; each non-affinity component $\Gamma$ determined through the quadratic fit is thus a function of $\varepsilon_{\rm cut}$.

\begin{figure}[h!]
    \centering
    \includegraphics[width=\textwidth]{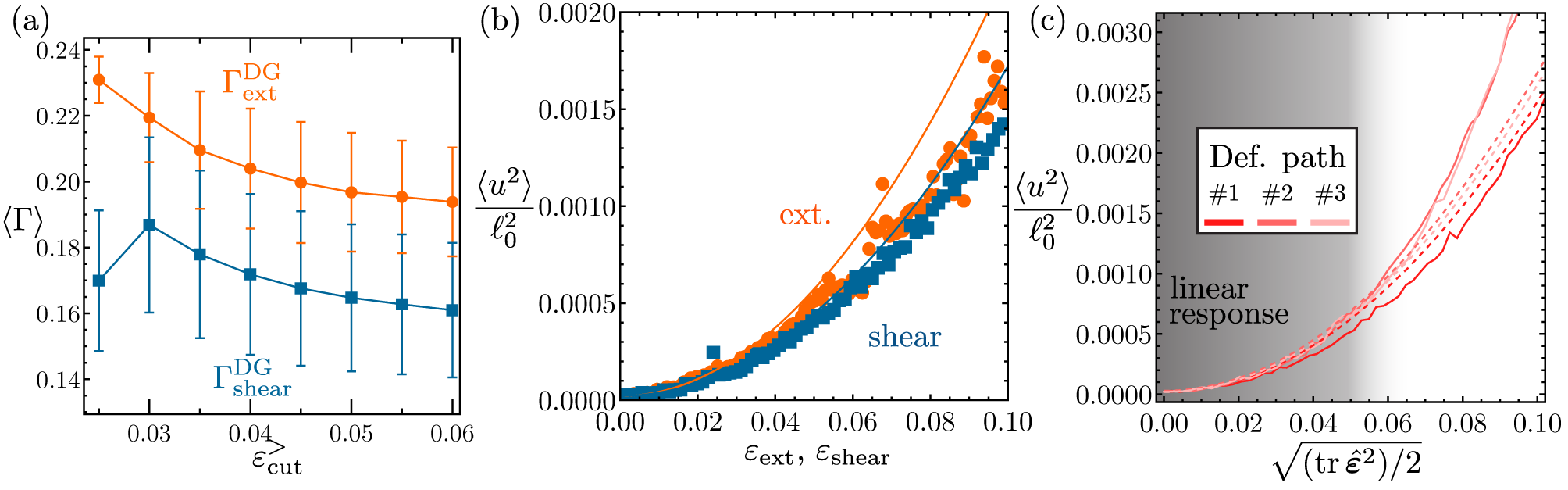}
    \caption{\label{fig:fitting} (a) Non-affinity measurements, as characterized by the mean $\langle\Gamma\rangle$ and standard deviation of the fitting parameter $\Gamma(\varepsilon_{\rm cut})$, where the cut-off strain for the fit, $\varepsilon_{\rm cut}$ is sampled over the interval $[0.02, \varepsilon_{\rm cut}^>]$, as a function of the upper limit of the cut-off strain, $\varepsilon_{\rm cut}^>$. Non-affinity parameter fit statistics are shown in orange for extensional strain and blue for shear strain. (b) Non-affine response $\langle |\mathbf{u}|^2 \rangle/\ell_0^2$ measured along the simulated extensional (orange) and shear deformation paths (blue) are shown from SCF calculations (points) and from the mean quadratic fit (solid curves). (c) The non-affine response of the three SCF-simulated deformation paths passing through the three triclinic crystal grains from experimental data are shown as a function of total strain $\sqrt{({\rm tr}\,\hat{\bm{\varepsilon}}^2)/2}$ (solid curves), focusing on the low-strain regime. By comparison, linear response predictions obtained from weighted combinations of response to extensional and shear deformation paths are shown as dashed curves. Deviation of SCF results from the linear response theory suggest that the linear regime is valid for strains up to $\lesssim 5\%$.}
\end{figure}

Given that the non-affinity measurement depends on the choice of $\varepsilon_{\rm cut}$, our accepted value involves a degree of error from this choice.
To estimate this error, we consider the variation of $\Gamma(\varepsilon_{\rm cut})$ over a range of $\varepsilon_{\rm cut}$ values.
This range is given by the interval $\left[\varepsilon^<_{\rm cut},\varepsilon^>_{\rm cut}\right]$; the lower limit, $\varepsilon^<_{\rm cut}$, is chosen to include enough data to provide a reasonable parabolic fit, given the noise present in SCF calculations; we choose $\varepsilon^<_{\rm cut} = 0.02$.
We calculate $\Gamma(\varepsilon_{\rm cut})$ over this range ($\varepsilon_{\rm cut} \in \left[\varepsilon^<_{\rm cut},\varepsilon^>_{\rm cut}\right]$); this function is made continuous using linear interpolation.
Next, we consider the statistics of $\Gamma(\varepsilon_{\rm cut})$ via the mean and variance,
\begin{equation}
    \begin{split}
        \langle \Gamma \rangle &\equiv \frac{1}{\varepsilon^>_{\rm cut} - \varepsilon^<_{\rm cut}}\int_{\varepsilon^<_{\rm cut}}^{\varepsilon^>_{\rm cut}} {\rm d}\varepsilon_{\rm cut}\, \Gamma(\varepsilon_{\rm cut}) \\
        \langle \delta\Gamma^2 \rangle &\equiv \frac{1}{\varepsilon^>_{\rm cut} - \varepsilon^<_{\rm cut}}\int_{\varepsilon^<_{\rm cut}}^{\varepsilon^>_{\rm cut}} {\rm d}\varepsilon_{\rm cut}\, \Gamma^2(\varepsilon_{\rm cut}) - \langle \Gamma \rangle^2
    \end{split}
\end{equation}
Using the variance in the fitting parameter as a measure of sensitivity to choice of $\varepsilon_{\rm cut}$, we consider the choice of $\varepsilon_{\rm cut}^>$ that minimizes the error $\sqrt{\vert\langle \delta\Gamma^2 \rangle\vert}$ (at fixed $\varepsilon_{\rm cut}^< = 0.02$); the mean fit $\langle \Gamma \rangle$ and error are shown as a function of $\varepsilon_{\rm cut}^>$ in Fig.~\ref{fig:fitting}(a).
For DG, we find that $\varepsilon_{\rm cut}^> \approx 0.04$ seems to be a reasonable upper limit on the fitting range, with values of the quadratic fitting parameter $\Gamma^{\rm DG}_{\rm shear}$ saturating beyond this point.
We adopt the mean values of the fitting parameter over this interval, $\langle \Gamma \rangle$, for both data sets as the accepted results, leading to estimates of the non-affinity of $\Gamma^{\rm DG}_{\rm ext} \approx 0.19 \pm 0.01$ and $\Gamma^{\rm DG}_{\rm shear} \approx 0.15 \pm 0.01$.
As shown in Fig.~\ref{fig:fitting}(b), these accepted values seem to fit the data quite well over a broad range of $\varepsilon \lesssim 0.1$.

\subsection{Estimating departure from linear response}

Using symmetry arguments, we showed that the linear response can be represented in terms of a linear combination of extensional $\varepsilon_{\rm ext}$ and shear $\varepsilon_{\rm shear}$ strains.
However, departures from this approximation are expected (and seen) to occur for larger deformations.
We can attribute departures from the linear theory presented here to several factors:
\begin{itemize}
    \item Higher-order corrections to the fixed-volume constraint ${\rm tr}\,\bm{\varepsilon} = 0 + \mathcal{O}({\rm tr}\, \bm{\varepsilon}^2)$ (see e.g.~\cite{WarnerTerentjev2007}).
    \item Increasing importance of higher-order strain couplings (e.g.~cubic terms of the form $\Gamma_{(3)}^{ijklmn}\varepsilon_{ij}\varepsilon_{kl}\varepsilon_{mn}$) resulting in ``mixing'' between strain terms that prevents factoring into powers of $\varepsilon_{\rm ext}$ and $\varepsilon_{\rm shear}$.
    \item Dilational modes (i.e.~${\rm tr}\,\bm{\varepsilon}$) in SCF calculations and experiments that, upon relaxation to thermodynamic equilibrium, spontaneously break network symmetry (i.e.~those symmetries ensured by space group of the original network), leading to non-affine deformations. 
\end{itemize}
The first two items listed above require a broader non-linear response extension to the work presented here.
The third item would represent a breakdown of a key assumption, namely the stability of network phases (with corresponding space group symmetries intact) under uniform dilations of space.
Non-affine response under dilations can be potentially incorporated into a mechanical network model by including forces that scale differently under re-scaling transformations of space; these may, for example, lead to ``buckling'' deformations of the network (sometimes seen in stiff filament networks~\cite{Conti2009}).
In our example system of block copolymer melts, the assumed incompressibility of the melt requires that dilations be accompanied by a flux of molecules into/out of the network, breaking the assumed fixed mass constraint.

To probe the extent of the linear response regime, we examine the deviation of SCF calculations from the expectations of our linear response theory.
Equilibrium calculations of the extensional and shear strain deformations of DG predict that the non-affine response to extensional strain is larger than the response to shear strain, $\Gamma_{\rm ext}^{\rm DG} > \Gamma_{\rm shear}^{\rm DG}$.
For all other deformation paths, we find it useful to turn to the ``polar projection'' of the deformation paths, characterized by their radius $\sqrt{({\rm tr}\,\hat{\bm{\varepsilon}}^2)/2}$ and angle $\alpha$, given by $\tan\alpha = \varepsilon_{\rm shear}/\varepsilon_{\rm ext}$.
For this analysis, we compute $\alpha$ from the first non-zero strain points along each deformation path.
For the deformation paths corresponding to the three DG deformation paths (labeled $\#1$, $\#2$, and $\#3$), we find $\alpha_1 \approx 66^\circ$, $\alpha_2 \approx 49^\circ$, and $\alpha_3 \approx 58^\circ$.
In these polar coordinates, our linear response theory predicts that the non-affine response of each of these paths can be computed from the extremal paths via
\begin{equation}
    \frac{\langle |\mathbf{u}|^2 \rangle}{\ell_0^2} = \left(\Gamma^{\rm DG}_{\rm ext}\cos^2\alpha + \Gamma^{\rm DG}_{\rm shear}\sin^2\alpha \right) \frac{{\rm tr}\,\hat{\bm{\varepsilon}}^2}{2} \equiv \Gamma^{\rm DG}_\alpha\,\frac{{\rm tr}\,\hat{\bm{\varepsilon}}^2}{2} \, ,
\end{equation}
where the polar angle $\alpha$ weights the two non-affinity parameters in the calculation of the non-affinity of for each path.
Since $0^\circ < \alpha_2 < \alpha_3 < \alpha_1 < 90^\circ$, as shown in Fig.~\ref{fig:def_paths}, we expect that $\Gamma^{\rm DG}_2 > \Gamma^{\rm DG}_3 > \Gamma^{\rm DG}_1$.
As shown in Fig.~\ref{fig:fitting}(c), this is indeed the case for small values of $\sqrt{({\rm tr}\,\hat{\bm{\varepsilon}}^2)/2}$.
Around $\sqrt{({\rm tr}\,\hat{\bm{\varepsilon}}^2)/2} \approx 0.05$, path $\#3$ starts overtaking path $\#2$ in scale of non-affine response, fully overtaking path $\#2$ around $\sqrt{({\rm tr}\,\hat{\bm{\varepsilon}}^2)/2} \lesssim 0.1$.
Moreover, the predicted linear, non-affine response shown by the dashed curves in Fig.~\ref{fig:fitting}(c), seems to follow the SCF data well for $\sqrt{({\rm tr}\,\hat{\bm{\varepsilon}}^2)/2} \approx 0.05$.
Thus, we estimate that the linear response regime is well-characterized for strain values $\varepsilon \lesssim 5\%$, with increasing non-linear response for higher strain values; as shown in Fig.~3(c) of the main text, non-linear effects rapidly become important for $\varepsilon \gtrsim 10\%$.

\section{Tomographic imaging of PS-PDMS}\label{sec:analysis}

The polystyrene-b-polydimethylsiloxane (PS-PDMS) diblock copolymer was synthesized by sequential anionic polymerization of styrene and hexamethylcyclotrisiloxane.
For the DG samples, the average molecular weight is 43.5 $\text{kg mol}^{-1}$ for PS and 29.0 $\text{kg mol}^{-1}$ for PDMS, with an overall composition of 40\% PDMS (by volume) and a molecular weight polydispersity index of 1.04.
The sample studied was cast slowly (over the course of one week) from a 10 wt\% solution (2 ml) in toluene. 
After drying, the sample was heated to $60^\circ$C for 3 days in a vacuum in order to remove any residual solvent.
For the DD samples, the average molecular weight is 51 $\text{kg mol}^{-1}$ for PS and 35 $\text{kg mol}^{-1}$ for PDMS, with an overall composition of 42\% PDMS (by volume) and a molecular weight polydispersity index of 1.05.
The sample studied was slowly cast over 2 weeks from a 5 wt \% solution (2 mL) in chloroform without further thermal annealing.

Given that the samples were heated to a temperature less than the glass temperature for polystyrene, the experiments reflect a structure that is not completely relaxed.
This may account for some of the discrepancy between the experimental data and SCF results and calls for a more in-depth study of structural relaxation under different processing conditions, including annealing protocols.

Data acquisition was performed using ``slice-and-view'' scanning electron microscopy (SVSEM), as detailed in ref.~\cite{Feng2019}.
A gallium-ion (Ga\textsuperscript{+}) beam was used to mill each sample surface.
Imaging of the free surface was performed using a 1-keV electron beam, enabling 2D density maps of PDMS relative to PS with a planar resolution of $\sim 3 {\rm nm}/{\rm pixel}$.
After imaging the free surface, a new sample surface was milled, resulting in a z-stack of 2D SEM images, where each z-slice was $\sim 3 {\rm nm}$ thick.
Details of the location of the reconstructions are provided upon reasonable request.
3D reconstruction of the density fields and skeletal graphs were then performed, following procedures in refs.~\cite{Prasad2018,Feng2019}.
In brief, the 3D density fields obtained from SVSEM were binarized based on block volume fraction of the sample and smoothed using a Gaussian filter.
Initial skeletonization was performed using a watershed algorithm; extraction of the skeletal network involved iterative merging of voxels in the skeletal set, until graphs with straight edges and 3- and 4-coordinated nodes were obtained.
Finally, a graph optimization was performed, wherein node positions were varied until a cost functional, given by the integral of the PDMS volume fraction along every strut in the graph was maximized.
This helps ensure that the skeletal graph is a good representation of the ``center'' of the PDMS domains.

We additionally employed the method outlined in ref.~\cite{Feng2019} to extract Bragg peaks from the 3D Fourier transform of the density field data.
Since deformed crystals transform the positions of the expected Bragg peaks of the corresponding cubic crystal, we used these data to extract affine deformation matrices $\bm{\Lambda}$ for each of the samples.
As $\bm{\Lambda}$ extracted in this way incorporates isometries of Euclidean space, which are irrelevant for describing deformations of the unit cell, we use a QR-decomposition to extract the triangular matrix part of the affine deformation, allowing for representation of $\bm{\Lambda}$ in the form given by Eq.~\ref{eq:def_mat}.
The extracted deformation matrices for each of the samples is listed below:
\begin{align}
\bm{\Lambda}^{\rm DG}_1 &= \left(\begin{array}{ccc}
1.033 & 0 & 0 \\
0.035 & 1.065 & 0 \\
-0.079 & 0.249 & 0.909
\end{array}\right) \\
\bm{\Lambda}^{\rm DG}_2 &= \left(\begin{array}{ccc}
1.071 & 0 & 0 \\
0.201 & 0.852 & 0 \\
0.018 & 0.091 & 1.097
\end{array}\right) \\
\bm{\Lambda}^{\rm DG}_3 &= \left(\begin{array}{ccc}
1.015 & 0 & 0 \\
-0.119 & 1.137 & 0 \\
0.266 & -0.056 & 0.866
\end{array}\right) \\
\bm{\Lambda}^{\rm DD}_1 &= \left(\begin{array}{ccc}
1.041 & 0 & 0 \\
-0.078 & 0.929 & 0 \\
-0.023 & -0.003 & 1.034
\end{array}\right) \\
\bm{\Lambda}^{\rm DD}_2 &= \left(\begin{array}{ccc}
0.994 & 0 & 0 \\
-0.006 & 1.165 & 0 \\
-0.108 & 0.113 & 0.863
\end{array}\right)
\end{align}

In order to extract the non-affine displacement, we used the same fitting method employed in SCF calculations.
Since the 3D-reconstructions of each grain contains several unit cells, each unit cell was considered an independent measurement of non-affine displacement, allowing for statistical estimates of error on each non-affinity measurement.
We measured the non-affine displacements of each node and averaged over identical nodes in adjacent unit cells to obtain a single-node average non-affine displacement measure $\overline{\left|\mathbf{u}^{(m)}\right|^2}$ for node $m$, as well as a corresponding standard deviation for the single node, $\sigma^{(m)}$.
The non-affine displacement averaged over a unit cell, $\langle |\mathbf{u}|^2 \rangle$, was then found by averaging over $\overline{\left|\mathbf{u}^{(m)}\right|^2}$, weighted by each node's individual standard deviation, i.e.
\begin{equation}
    \begin{split}
        \langle |\mathbf{u}|^2 \rangle &= \frac{1}{\sigma_{\rm tot}^2}\sum_{m}\frac{\overline{\left|\mathbf{u}^{(m)}\right|^2}}{(\sigma^{(m)})^2} \\
        \sigma^{-2}_{\rm tot} &= \sum_m (\sigma^{(m)})^{-2}
    \end{split}
\end{equation}
where $\sigma_{\rm tot}$ is the overall standard deviation of the averaged measurements.
Note that this same method of extracting data from multiple unit cells within each sample was used to calculate error on strut length variance and node volume variance.

\subsection{Comparison between DD experiments, SCF computations, and LNT}

\begin{figure}[h!]
    \centering
    \includegraphics[width=\textwidth]{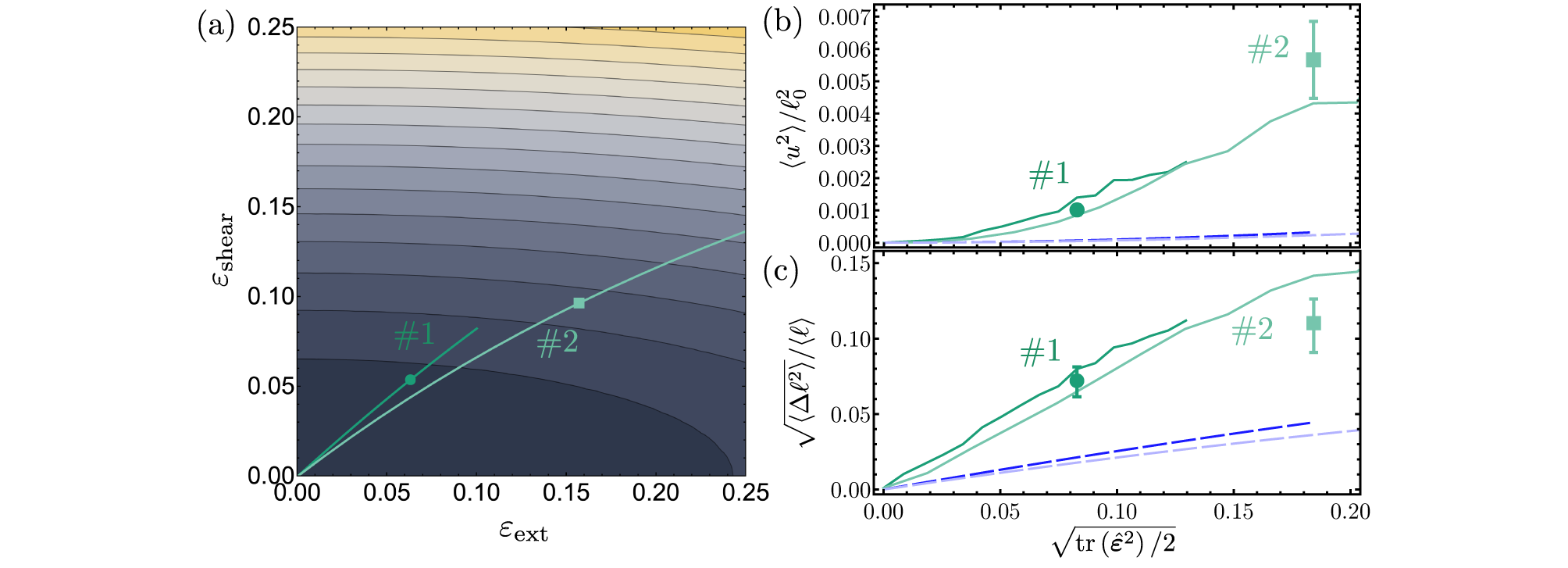}
    \caption{\label{fig:DD_results} (a) Deformation paths \#1 and \#2 for DD (solid curves), projected onto the space spanned by the linear combination of extensional strain $\varepsilon_{\rm ext}$ and shear strain $\varepsilon_{\rm shear}$. These curves pass through points representing the two triclinic DD grains from experimental data. (b) Non-affine displacement response $\langle |\mathbf{u}|^2 \rangle/\ell_0^2$ of experimental data (points), SCF calculations (curves), and LNT predictions (dashed curves). (c) r.m.s.~strut length variations for the same data sets.}
\end{figure}

Here, we report comparisons between measurements of DD grains, simulated deformation paths in SCF, and predictions from LNT.
The results are summarized in Fig.~\ref{fig:DD_results}.
As shown in Fig.~\ref{fig:DD_results}(a), while the two grains have significantly different deformations, their respective simulated deformation paths have very similar trajectories in the linear regime, making it difficult to differentiate their linear response behavior.
Indeed, both the SCF and LNT predictions show nearly-overlapping results for the non-affine displacement $\langle |\mathbf{u}|^2 \rangle/\ell_0^2$ (shown in Fig.~\ref{fig:DD_results}(b)) and the r.m.s.~length variations $\sqrt{\langle \Delta \ell^2\rangle}/\langle\ell\rangle$ (shown in Fig.~\ref{fig:DD_results}(c)).
Note that for sample \#1, we observed a slight gradient to the lattice spacing, so to minimize the effect of this gradient on the calculation of the non-affine displacement, we excluded points a distance greater than $2\ell_0$ from the center of the 3D reconstruction; inclusion of further points increased the measured value of $\langle |\mathbf{u}|^2 \rangle$ by an order of magnitude.
Moreover, due to relatively large number of 3-coordinated nodes observed in sample \#2, we set the cutoff on minimum graph edge length during the reconstruction to be large enough to yield only 4-coordinated nodes.
As discussed in Sec.~\ref{sec:node_splitting} below, deformations of DD, particularly those that lead to large extensional strain, result in splitting of 4-coordinated nodes into pairs of 3-coordinated nodes.
Since sample \#2 has a considerably larger value of $\varepsilon_{\rm ext}$ than sample \#1, the increase in node splitting is consistent with this observation.

\section{Nodal volume measurements in DG}

Calculations of nodal volume for 3D reconstructions of experiments and SCF calculations were performed by associating each voxel in the 3D A-block density field to the nearest node, according to Euclidean distance.
Given the volume $\delta V$ of each voxel and local A-block volume fraction $\phi_v$ at voxel $v$, the total A-block volume $V^{(m)}$ associated to node $m$ is given by
\begin{equation}
    V^{(m)} = \delta V\,\sum_{v \in \mathcal{N}_m} \phi_v \, ,
\end{equation}
where $\mathcal{N}_m$ is the set of voxels associated to node $m$.
For the experimental measurements, we only calculated the volumes of nodes that are a suitable distance away from the sample boundary, to reduce error.
This amounted to nodes that lie close enough to the center of the data set; we found that only calculating nodal volume for nodes within a distance of $4\ell_0$ of the center was sufficient for both reducing boundary error and obtaining a suitable number of corresponding nodes to obtain error estimates via the method outlined in Sec.~\ref{sec:analysis}.

For LNT comparisons, we modeled the volume of each node as the volume of the Voronoi cell generated by the node.
In the cubic configuration, all nodes are equivalent, so each Voronoi cell is equivalent.
In each deformed configuration, node displacements lead to variations in Voronoi cell volumes.
We find that the r.m.s.~volume variance computed in this way is consistently less than the volume variance computed from SCF and experiment.
This, in part, is due to variations in network domain thickness that are not captured in the Voronoi model.
Moreover, the polygonal faces of Voronoi cells only roughly approximate the domain boundaries of network phases, which are better approximated as minimal surfaces (see e.g.~\cite{Reddy2022,Dimitriyev2023}).

\begin{figure}[h!]
    \centering
    \includegraphics[width=.8\textwidth]{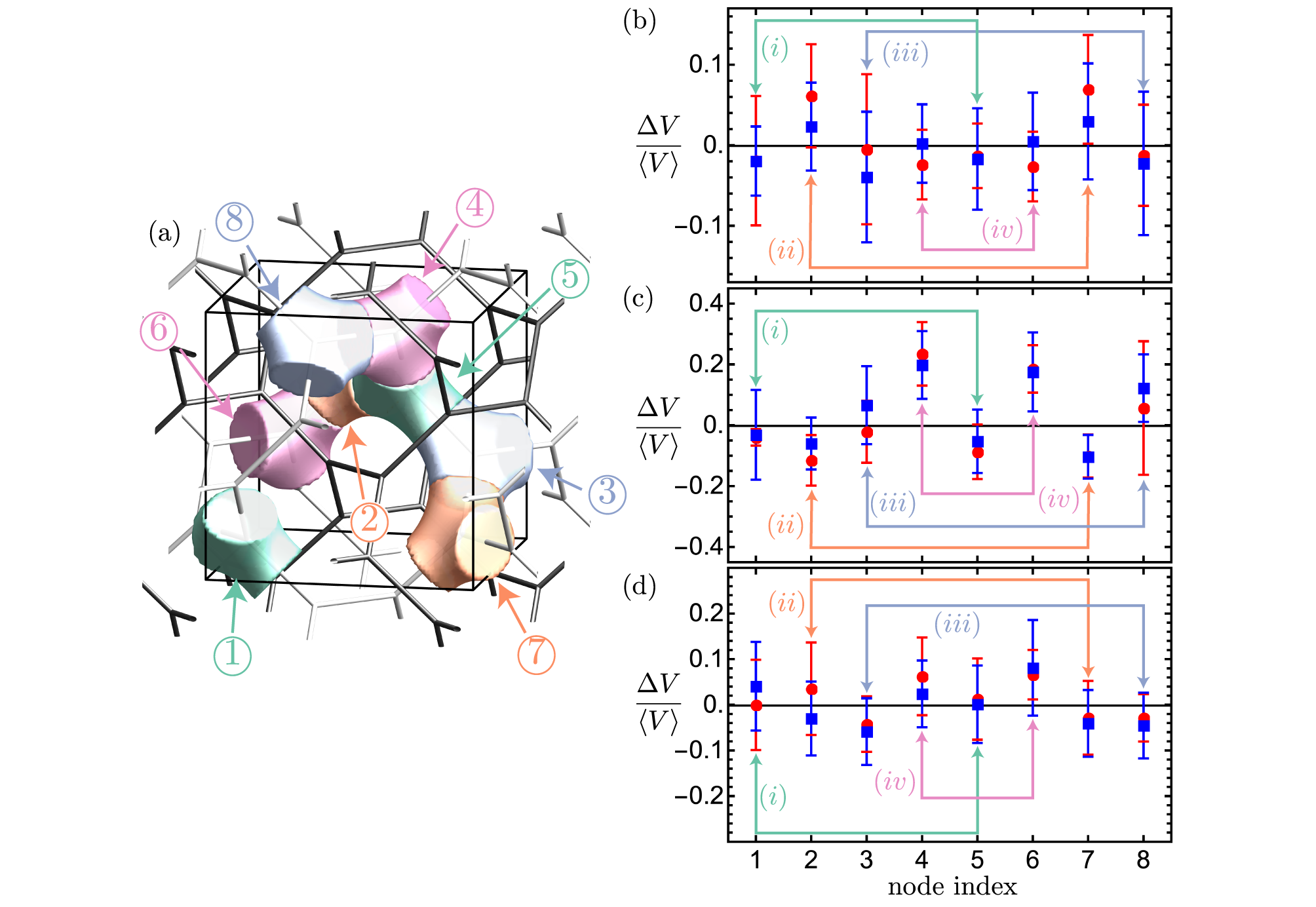}
    \caption{\label{fig:DG_node_volumes} (a) Single DG skeletal graph showing the 8 nodal regions and pairs of equivalent nodes represented by the same color. (b)-(d) Volume variation with respect to the mean $\Delta V/\langle V\rangle$ for DG grains \#1 - \#3, where red and blue data points represent contributions from the two separate networks. Subsets of identical nodes are labeled by \textit{(i)-(iv)}.}
\end{figure}

\subsection{Nodal polydispersity in DG}

While all nodal regions within an undeformed, cubic DG, DD, or DP crystal are equivalent and can be identified as ``mesoatomic'' building blocks, symmetry-breaking deformations can result in different deformations amongst the various nodes.
Our analysis of the trihedral graphs, as reported in Eq.~\ref{eq:tri_result}, shows that the response of a nodal region depends on the orientation of the node.
In the DG skeletal graphs, each graph has 8 nodes, and of those nodes, pairs share the same orientation, as shown in Fig.~\ref{fig:DG_node_volumes}(a).
These orientation-sharing nodes additionally have the same neighbors: they are identical under a half-translation of the unit cell (i.e., by $\left(1,1,1\right)/2$).
Since the second skeletal graph is obtained by a translation and inversion of the first, it shares the same node orientations; thus, out of 16 nodes, there are 4 equivalent subgroups.
Moreover, since the triclinic deformations considered here preserve inversion symmetry, these 4 nodes subsets should be preserved in triclinic DG networks.
One metric of nodal geometry is provided by the nodal volume; variance in nodal volume indicates some degree of deformation (note that nodes may deform without changing volume, however).
The variation in nodal volume amongst the 8 different nodes from each skeletal graph (denoted by red and blue points) obtained from the 3 experimental data sets is shown in Fig.~\ref{fig:DG_node_volumes}(b)-(d).
We see that patterns in the range of volume variations for each of the 8 indexed nodes consistently fall into 4 equivalent nodes between the two networks, labeled \textit{(i)-(iv)}.
We find that the two networks in each data set are nearly identical according to volume variance, with slight differences, notably for sample \#2. 
While further studies are needed to better quantify nodal shape variations, we can nonetheless identify persistent sets of identical nodes within DG networks.
By contrast, DD all nodes are equivalent under translation and inversion and do not have this same mode of nodal polydispersity.
Instead, we find that DD nodes can undergo ``splitting'' into pairs of 3-coordinated nodes, as described in Sec.~\ref{sec:node_splitting}.

\section{Node splitting in double diamond (DD)}\label{sec:node_splitting}

\begin{figure}[h!]
    \centering
    \includegraphics[width=.8\textwidth]{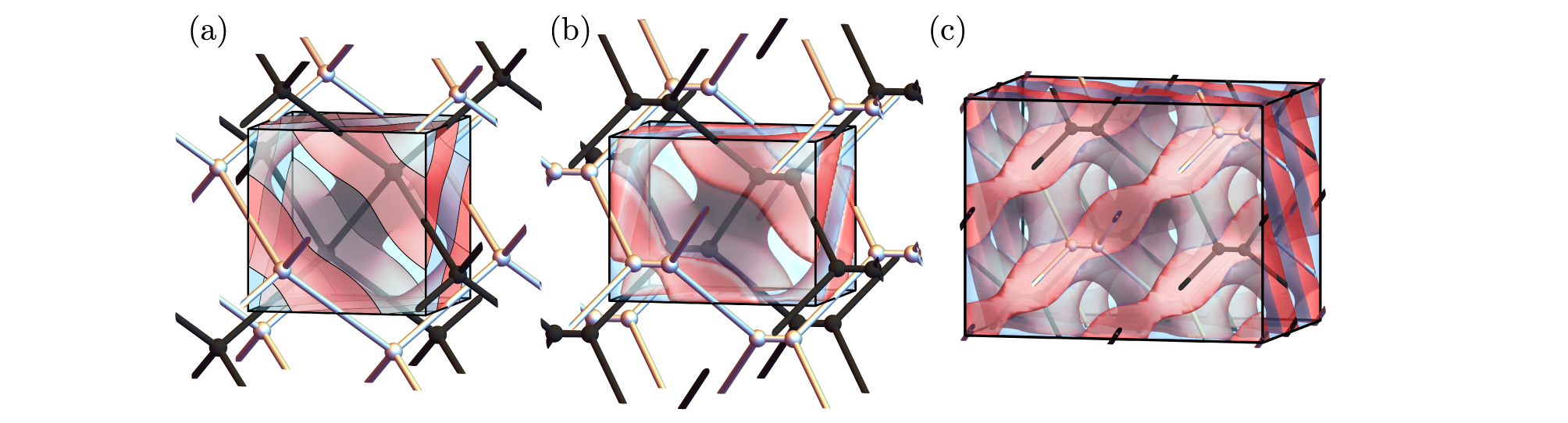}
    \caption{\label{fig:DD_node_fission} (a) DD cubic unit cell from SCF calculations, along with skeletal graphs. (b) Relaxed DD structure after imposed tetragonal deformation ($\varepsilon_{\rm ext} \approx 16\%$), showing that the 4-coordinated split into pairs of 3-coordinated nodes. (c) View of 8 unit cells of the tetragonal-DD network, showing the full structure of the double-network.}
\end{figure}

In order to calculate the non-affinity of the deformed DD networks, depicted in Fig.~\ref{fig:DD_node_fission}(a), we imposed a constraint on the edge length cutoff for the skeletonization algorithm that ensured the skeleton graphs consist of 4-coordinated nodes.
However, if we relax the edge cutoff to standard values used for constructing DG networks, we find that above a modest value of extensional strain, $\varepsilon_{\rm ext} \gtrsim 5\%$, the 4-coordinated nodes split into pairs of 3-coordinated nodes.
As the strain is increased, the length of the strut joining the new nodes increases, eventually reaching a length that is about $20\%$ of the initial strut length of the DD near strain values of $\varepsilon_{\rm ext} \sim 16\%$, depiced in Fig.~\ref{fig:DD_node_fission}(b),(c).
Since the new nodes are close to planar and have strut angles close to $120^\circ$, we expect that such networks are well-described by LNT.
As the equilibrium networks of LNT minimize their total length at constant volume, this suggests that such ``topological transitions'' of the graph structure can be rationalized by the mathematics of length-minimizing, ``Steiner'' networks.
Recent work~\cite{Alex2023} has shown that the skeletal network corresponding to the DG, the \texttt{srs} network (named for the strontium silicide crystal structure), is the globally length-minimizing, triply-periodic Steiner network.
We recognize that the skeletal networks of the tetragonally-deformed DD closely resemble the \texttt{ths} network (named for the thorium silicide crystal structure), which include Steiner networks with lengths slightly larger than the \texttt{srs}.
The \texttt{ths} networks, also called (10,3)-b~\cite{Wells1977}, consist of planar ``zig-zag'' strut patterns that resemble the skeletal structure of the O\textsuperscript{70} (orthorhombic $Fddd$ phase)~\cite{Bailey2002,Tyler2005}.
Interestingly, the SCF calculations showed a continuous, monotonic free-energy \emph{decrease} from the cubic state, suggesting that under conditions of fixed volume (i.e.~fixed total mass), the cubic DD is weakly unstable to tetragonal deformations; at $\varepsilon \approx 16\%$, we found a slight free-energy decrease of $\sim -0.1\%$ relative to the cubic-DD.
SCF calculations starting for the tetragonal-DD and deforming back to the cubic-DD showed the same free-energy change, indicating that this transition is indeed continuous.
We leave the exploration of the node-splitting mechanics and thermodynamics for future study.

\bibliography{refs}